\documentclass[journal,onecolumn,12pt]{IEEEtran}

\newif\ifcomen
\comentrue

\usepackage{tikz}
\usepackage{caption}
\usepackage[utf8]{inputenc}
\usepackage{graphicx}
\usepackage{esvect}
\usepackage{comment}
\usepackage{cite}
\usepackage{epstopdf}
\usepackage{amsthm}
\usepackage{array}
\usepackage{float}
\usepackage{setspace}
\usepackage{algpascal}

\usepackage{amssymb}
\usepackage{microtype}
\usepackage{color}
\usepackage{relsize}
\usepackage{multirow}
\usepackage{balance}
\usepackage{amsmath}
\usepackage{hhline}
\usepackage{flushend}
\usepackage{cancel}
\usepackage{enumitem}
\usetikzlibrary{positioning,arrows,automata}
\usepackage{footnote}
\usepackage{footmisc}
\usepackage{mathtools}
\DeclarePairedDelimiter{\ceil}{\lceil}{\rceil}

\newtheorem{prop}{Proposition}

\usepackage{algorithm,algpseudocode}
\algnewcommand\algorithmicforeach{\textbf{for each}}
\algdef{S}[FOR]{ForEach}[1]{\algorithmicforeach\ #1\ \algorithmicdo}

\algnewcommand\algorithmicForRange{\textbf{for} $k \gets 1$ to}
\algdef{S}[FOR]{ForRange}[1]{\algorithmicForRange\ #1\ \algorithmicdo}

\usetikzlibrary{arrows,shapes,snakes,automata,backgrounds,petri}

\definecolor{DarkGreen}{rgb}{0.1,0.5,0.1}
\definecolor{DarkRed}{rgb}{0.5,0.1,0.1}
\definecolor{DarkBlue}{rgb}{0.1,0.1,0.5}
\definecolor{DarkPurple}{rgb}{0.5,0.2,0.5}
\definecolor{DarkTurquoise}{rgb}{0.1,0.5,0.5}

\newcommand{\off}[1]{}

\algdef{SE}{For}{End}[2]{
  \textkeyword{for} \(#1\) \textkeyword{to} \(#2\) \textkeyword{do begin}}{%
  \textkeyword{end}}

\title{Bringing Network Coding into SDN: A Case-study for Highly Meshed Heterogeneous Communications\thanks{ A. Cohen, H. Esfahanizadeh,  and M. M\'{e}dard are with the Research Laboratory of Electronics, MIT, Cambridge, MA, USA (e-mail:cohenale@mit.edu; homaesf@mit.edu; medard@mit.edu). B. Sousa is with the Department of Informatics Engineering of the University of Coimbra, Portugal (e-mail:bmsousa@dei.uc.pt). J. P. Vilela is with CRACS/INESCTEC and the Department of Computer Science of the Faculty of Sciences of the University of Porto, Portugal (e-mail:jvilela@fc.up.pt). M. Lu\'is is with Instituto Superior de Engenharia de Lisboa and Instituto de Telecomunica\c{c}\~oes, Portugal (e-mail:nmal@av.it.pt). D. Raposo is with Instituto de Telecomunica\c{c}\~oes, Portugal (e-mail:dmgraposo@av.it.pt).  F. Michel is with  UCLouvain, Belgium (e-mail:francois.michel@uclouvain.be). S. Sargento is with the University of Aveiro and Instituto de Telecomunica\c{c}\~oes, Portugal (e-mail:susana@ua.pt).\newline Part of this work is submitted to IEEE Communications Magazine on Transport Layer Innovations for Future Networks.}}
\author{Alejandro Cohen, Homa Esfahanizadeh,
Bruno Sousa, Jo\~{a}o P. Vilela, Miguel Luis,\\ Duarte Raposo, François Michel,
Susana Sargento, and Muriel M\'{e}dard}
\begin{document}

\maketitle

\begin{abstract}
Modern communications have moved away from point-to-point models to increasingly heterogeneous network models. In this article, we propose a novel controller-based protocol to deploy adaptive causal network coding in heterogeneous and highly-meshed communication networks. Specifically, we consider using Software-Defined-Network (SDN) as the main controller. We first present an architecture for the highly-meshed heterogeneous multi-source multi-destination networks that represents the practical communication networks encountered in the fifth generation of wireless networks (5G) and beyond.
Next, we present a promising solution to deploy network coding over the new architecture.
In fact, we investigate how to generalize adaptive and causal random linear network coding (AC-RLNC), proposed for multipath multi-hop (MP-MH) communication channels, to a protocol for the new multi-source multi-destination network architecture using controller. To this end, we present a modularized implementation of AC-RLNC solution where the modules work together in a distributed fashion and perform the AC-RLNC technology. We also present a new controller-based setting through which the network coding modules can communicate and can attain their required information.
Finally, we briefly discuss how the proposed architecture and network coding solution provide a good opportunity for future technologies, e.g., distributed coded computation and storage, mmWave communication environments, and innovative and efficient security features.
\end{abstract}

\section{Introduction}\label{sec_intro}

The increasing demand for network connectivity and high data rates requires the efficient utilization of all possible resources. In recent years, the connectivity moved forward from point-to-point schemes to multi-source multi-destination (MS-MD) meshed network of interconnected nodes in which intermediate nodes can cooperate and share the physical  resources for efficient and reliable communications.

The current state-of-the-art research lacks a comprehensive understanding of the interplay among coding and scheduling within a heterogeneous highly meshed MS-MD network context. To embrace the groundbreaking 5G network at massive scale in their dense environments, we present efficient adaptive and causal random linear network coding (AC-RLNC) techniques to provide robustness to service degradation and provide effective and reliable network communications. Furthermore, we will exploit the information that can be provided by the Software-Defined-Network (SDN), \cite{Rezende2019}. In the adaptive network coding, we suggest learning, in real time, the current erasure pattern and the network link rates, using a joint control, to consequently improve the decisions.

\begin{figure}
    \centering
    \includegraphics[trim=0cm 0.0cm 0cm 1cm,width = 0.7 \columnwidth]{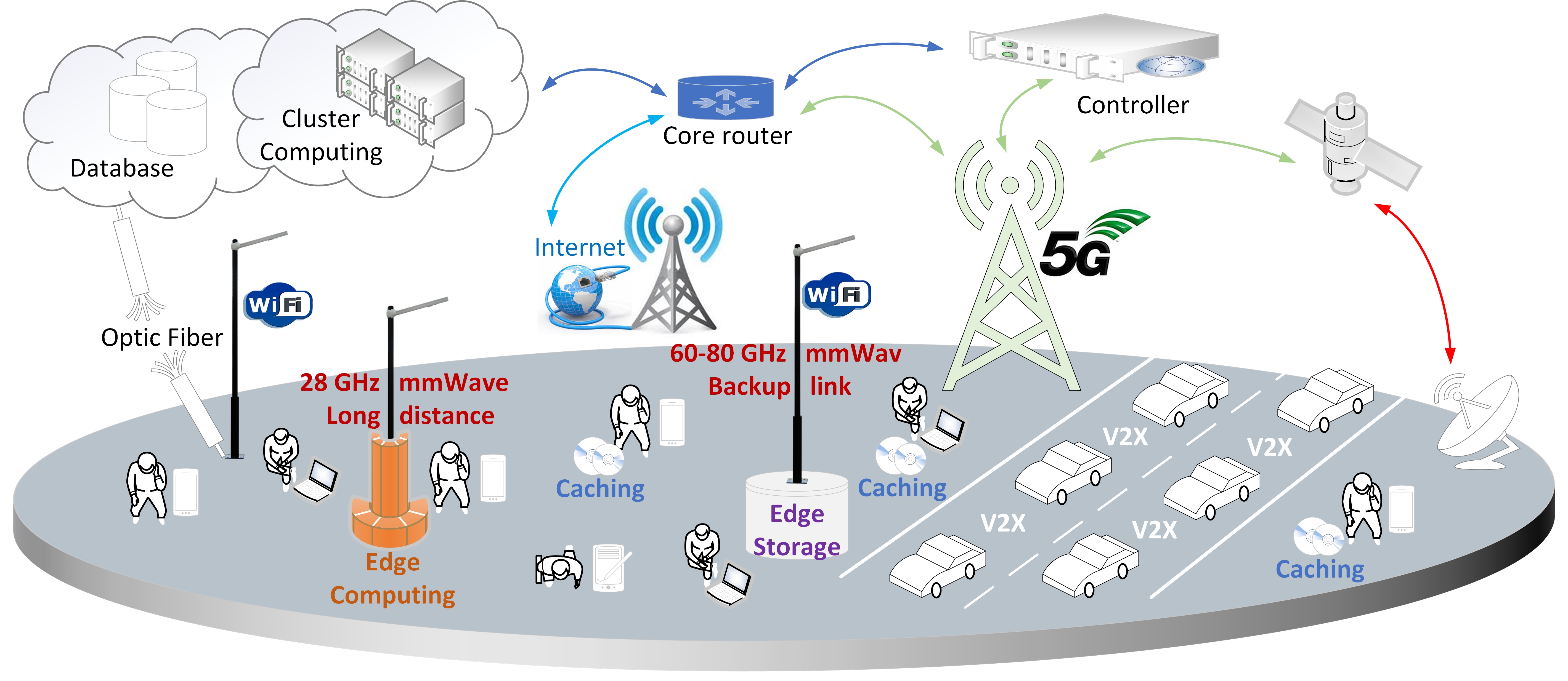}
    \caption{Heterogeneity in meshed communications.}
    \label{smart_city}
\end{figure}

The proposed network-coding based solution offer benefits for ultra-reliable communication for highly-interconnected MS-MD networks. As a promising example, the communication solution presented in this article can be very beneficial for realizing the advanced requirements of Smart Cities. One of the main priorities consists on the integration of heterogeneous communication infrastructures and processes to enable the right environment where digital networks and services can prosper, see Fig.~\ref{smart_city}.

Upcoming communication networks, 5G and beyond, will be faced with stringent capacity and latency requirements. With an impressive growth of new services and devices (consuming and producing huge amounts of data), mobile networks must become ready to grant the expected quality of service to its users. However, achieving the 5G key performance indicators especially in highly dense networks of devices, such as urban areas of the major cities, will require complementary solutions. For example, to cope with the lack of connectivity, or simply to increase the backhaul link capacity traditionally supported by fiber links, high frequency and high capacity millimeter-wave (mmWave) communications have proven to be an excellent candidate \cite{bib:pi2016mmwave}. In such situations, a multitude of end-to-end paths are provided, bringing multipath diversity to the network that could be explored by network coding techniques.

The AC-RLNC solution presented in this article is extremely helpful in realizing the throughput and delay requirements of the envisioned services in Smart Cities as considered recently in \cite{noauthor_snob-5g_nodate} for wireless backhaul solutions in 5G networks. The availability of multiple paths, using AC-RLNC to reach the desired throughput-delay tread-off, enhances the resilience of real-time applications like video streaming, vehicle to everything (V2X) communications, etc.

In this paper we propose an architecture that fosters the exchange of control information between network coding communication solution, SDN, network function virtualization (NFV) management, and orchestration components for general meshed heterogeneous communications and, as an propitious example, for services in Smart Cities powered by 5G wireless backhaul technologies.

The structure of this work is as follows. In Section \ref{sec:overview}, we provide an overview of classical communication solutions over MS-MD networks. In Section \ref{sec_background_ac_rlnc}, we present the necessary background on AC-RLNC for single-path (SP), multipath (MP), and multipath and multi-hop (MP-MH) networks. In Section \ref{sec_architecture}, we introduce the network architecture we propose for the heterogeneous MS-MD communications. In Sections \ref{sec:Controller} and \ref{sec:MS-MD-AC-RLNC}, we present our scheme for communication over meshed MS-MD networks. In particular, Section \ref{sec:Controller} introduces the proposed scalable, secure, and efficient SDN Controller (SSE-SDNC), while Section \ref{sec:MS-MD-AC-RLNC} introduces the generalized AC-RLNC in conjunction with the SSE-SDNC controller. In Section \ref{sec:visions}, we show a few important examples of enabling technologies and features. Table~\ref{table : definition} lists the symbols that we use throughout this paper, along with their definitions.

\begin{table}
\normalsize
    \centering
    \begin{tabular}{|l|l|}
        \hline
        \textbf{Param} & \textbf{Definition}\\\hline
        $P$, $H$& number of paths and number of hops \\\hline
        $\epsilon_{p,h}$ &  erasure probability of the $p$'th path and $h$'th hop \\\hline
        $r_{p,h}$& $1-\epsilon_{p,h}$, rate of the $p$'th path and $h$'th hop \\\hline
        $t$& time slot index\\\hline
        $M$ & number of information packets\\\hline
        $p_i$ & information packet, $i\in [1,M]$\\\hline
        $\mu_i \in \mathbb{F}_z$ & random coefficients \\\hline
        $c(t,p)$ & RLNC to transmit at time slot $t$ on $p$'th path\\\hline
        $k$ & number of packets in window, $RTT-1$ \\\hline
        $EW$ & end window of $k$ new packets \\\hline
        $w_\text{min}$,  $w_\text{max}$& limits of the effective window
        \\\hline
        $w$& length of the effective window \\\hline
        $r_{G_p}$& rate of $p$'th global path
        \\\hline
        $\bar{o}$ & maximum window size \\\hline
        $m_{G_p}$ & number of FECs for the $p$'th global path\\\hline
        $a_{d_g}$ & number of added DoFs\\\hline
        $m_{d_g}$ & number of missing DoFs\\\hline
        $th$ & re-transmission margin \\\hline
        $\Delta$ & $P\cdot(d-1-th)$, re-transmission parameter \\\hline
        $N_\text{new}$ & number of NEW-RLNC packets \\\hline
        $N_\text{ret}$ & number of REP-RLNC packets \\\hline
    \end{tabular}
    \caption{Symbol definitions}
    \label{table : definition}
\end{table}

\section{Overview of Multiple Source Multiple Destination Networks}\label{sec:overview}

\subsection{RLNC and MPTCP Communication}

Transmission Control Protocol (TCP) and Internet Protocol (IP) were first proposed in the '1970s \cite{kahn1974protocol,cerf1974specification}. The Classical TCP in the transport layer was first considered for a single point-to-point connection \cite{kozierok2005tcp,fall2011tcp}, and the protocol was not capable of efficiently supporting the multipath communications as it relates applications to source and destination IP addresses and ports. Moreover, even for a single communication with a lossy link or with varying Round Trip Time (RTT) delay, TCP is known to be sub-optimal with major limitation on loss recovery time \cite{martin2002tcp,johansson2016congestion}.

\begin{figure}
    \centering
    \includegraphics[width=0.6\textwidth]{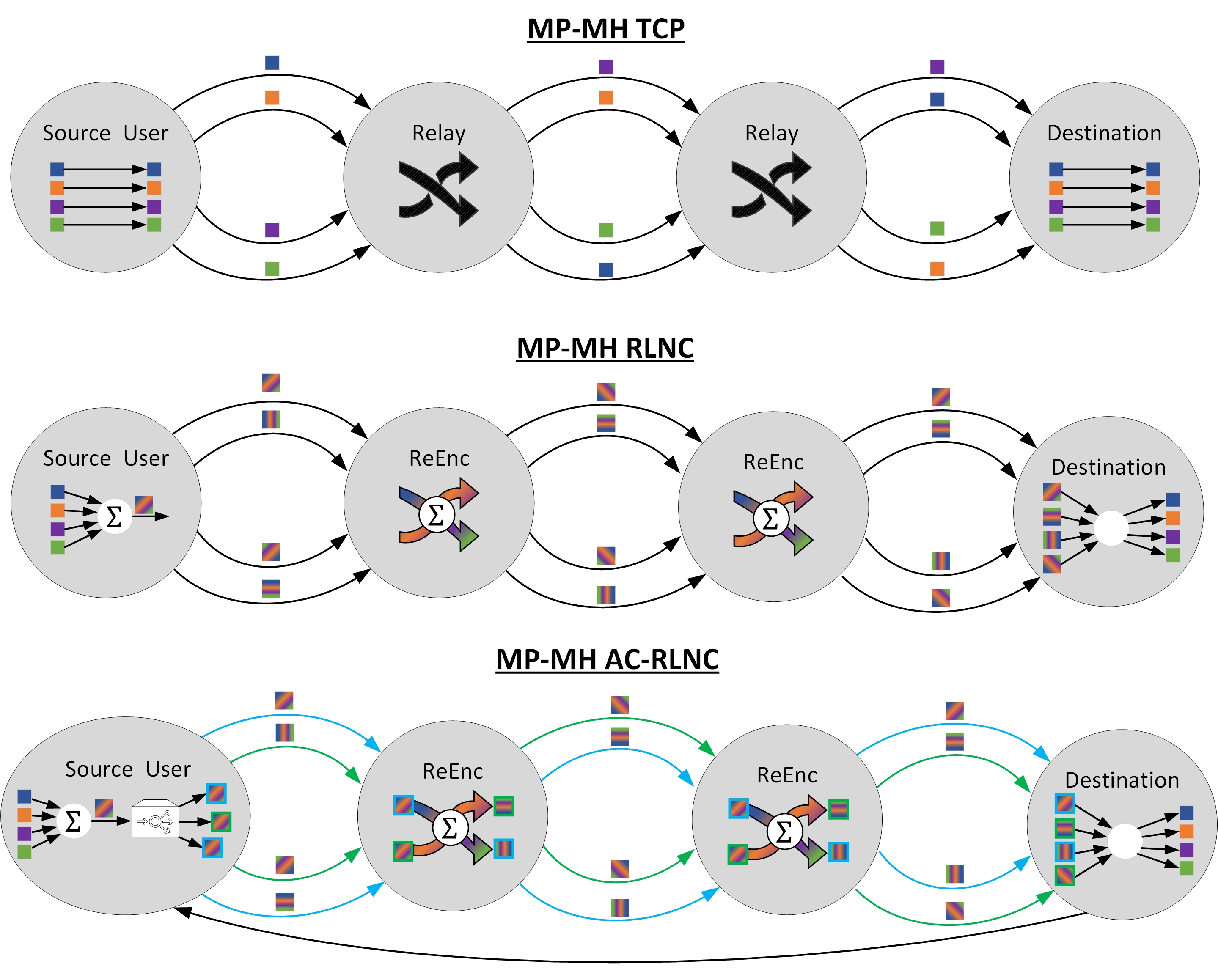}
    \caption{For a MP-MH communication network: Top panel depicts traditional MPTCP communication solution; Middle panel depicts RLNC communication solution; Bottom panel depicts AC-RLNC communication solution.
    \vspace{-0.4cm}}
    \label{fig:Classical_vs_RLNC3}
\end{figure}

Multipath TCP (MPTCP) has been considered in the last decade, allowing multipath communication over naive selection of IP connections\footnote{We refer to all the direct available connections between the source and the destination without considering the link rates and the bottleneck effects in the multi-hop communications as naive selection of global paths.} between a source and a destination, for a single application using multiple TCP subflows simultaneously~\cite{paasch2014multipath}. This multipath solution has been designed to be similar to regular TCP to cope with middleboxes.
Fig. \ref{fig:Classical_vs_RLNC3}, top panel, illustrates the MPTCP solution, by simultaneously using multiple independent TCP subflows, over the naive multipath connections between the source and the destination. Although MPTCP provides better utilization of the available resources in a multipath communication, delay and loss variances result in packet reordering following on specific lost packets, increased out-of-order buffer, reduced overall throughput, which causing MPTCP solution at times performs worse than traditional TCP \cite{yedugundla2016multi,ferlin2016blest}.

Random Linear Network Coding (RLNC), was first introduced in the '2000s, achieving the min-cut max-flow of the multipath and multi-hop (MP-MH) networks by mixing long blocks of data with random linear coefficient over a large enough field \cite{ho2006random}. The RLNC-coded packets transmitted across the multipath network are random linear combinations of the raw packets of information in a block \cite{schneuwly2020discrete}. The receiver, when obtains sufficient linear combinations, can decode all the information by performing a Gaussian elimination on a linear system \cite{chou2003practical,ho2006random,patterson2014and}. Consider a sequence of $M$ information packets $\{p_1,p_2,\dots\,p_M\}$ that need to be transmitted from a source to a destination through a network of interconnected nodes. Instead of sending the original packets, random linear combinations of these packets are transmitted, i.e., $\sum_{i=1}^{M}\mu_ip_i$ where the coefficients $\{\mu_1,\mu_2,\dots,\mu_M\}$ are randomly drawn at time step $t$.

Unlike MPTCP, to achieve the capacity of the MP-MH network, the raw data transmitted by the source (in a block) is also mixed and re-encoded at the intermediate nodes\footnote{In the re-encoding process proposed in the traditional RLNC, the intermediate nodes (so-called ReEnc) draw random coefficients for the packets coming from all the input links and prepare a new linear combination to transmit over all the output links. Note that decoding is not needed in the intermediate nodes.} \cite{ho2006random}. In a lossy multipath communication network with varying delay, the RLNC results in a robust solution in which the decoder, which decodes a block of linear combinations of the raw information, is not highly affected by the missing of specific packets. Fig. \ref{fig:Classical_vs_RLNC3}, middle panel, illustrates the RLNC solution mixing all the raw data at the source and re-mixing all the received data at the intermediate nodes using new random coefficients. Although RLNC in the MP-MH network may achieve the maximum throughput in the large blocklength regime, emerging advanced applications demand low in-order delivery delays while the high data rates require all the available resources of the network. Traditional information-theoretic solutions, including RLNC that requires large blocklength \cite{luby1997practical}, are not able to reach this desired trade-off.

Recently, in single path communication using TCP and multipath communication using MPTCP, a priori forward error correction (FEC) according to the feedback acknowledgments was considered to reduce the in-order delay \cite{shokrollahi2006raptor,luby2002lt,GuoShiCaiMed2013,ferlin2018mptcp}. Coding techniques, e.g., MDS codes \cite{badr2016layered,fong2019low} and systematic block codes \cite{cloud2015coded}, were considered as well. In particular, to compensate the average erasures in the lossy channels, RLNC is employed by periodically sending a priori FEC re-transmissions, i.e., the RLNC-coded packets generated from raw information in a sliding window block that has been used before for generating coded packets \cite{sundararajan2011network,kim2012network,cloud2013multi}.
Coded TCP in single path proposes to send a different amount of repair symbols depending on the current loss rate. Although those solutions proposed recently in the literature are reactive to the feedback acknowledgments, namely causal, none of those solutions are tracking the varying channel condition and rate (not adaptive). This results in high delay or lower throughput that may be far from the desire reliability-delay trade-off.

\subsection{SDN and Multiple Paths Communication}
Multipath communication in conjunction with SDN has been used to enhance the performance and reliability in data centers and communication networks. When a node needs to communicate with another using MPTCP, through a multipath network, SDN can provide the information of the underlying infrastructure between the two nodes. The provided information (e.g. intermediate nodes) can be employed to determine the number of TCP connections that need to be created in order to leverage the availability of multipath network resources.
Indeed, the combination of SDN and MPTCP leads to an efficient network resource utilization and congestion reduction, since the usage of idle and overloaded links are decreased. In addition, it also overcomes the challenges associated with traditional multipath routing approaches resulting from the Equal Cost Multipath (ECMP) algorithm, which randomly selects a path for the load balancing~\cite{Lin2019}.

Other protocols, such as Stream Control Transmission Protocol (SCTP) and Quick UDP Internet Connections (QUIC) given in  \cite{langley2017quic,de2019pluginizing} can also leverage the benefits of employing SDN, which allows them to establish a mapping between the routing information and the application streams. In fact, SDN controller can manage the configuration of devices (e.g. using OpenFlow, P4) to assign the transmission of packets from each stream over different paths \cite{Rezende2019}.

SDN is also combined with segment routing approaches to decrease the delay in the transmission of packets and also to decrease the traffic volume in the distinct sub-networks~\cite{Pang2017}. Segment routing is a traffic engineering mechanisms that allows to use segment as an ordered list of instructions for packet routing.

\section{Overview of Adaptive and Casual Random Linear Network Coding (AC-RLNC)}\label{sec_background_ac_rlnc}

To achieve the desired delay-throughput trade-off demanded by the emerging advanced applications and technologies, i.e., low in-order delay and high data rate, recently proposed an Adaptive and Causal RLNC (AC-RLNC) solution for a communication with one source user and one destination over single path (SP) communication \cite{cohen2020adaptive} and over MP-MH network  \cite{cohen2020adaptiveMPMH}. In this section, we briefly review the AC-RLNC communication solution. AC-RLNC solution is adaptive to the link condition, and it is causal since the data transmissions from the source user depend on the particular erasure realizations, as reflected in the feedback acknowledgments from the destination. That is, AC-RLNC can track the erasure pattern of the links in the network and adaptively adjust re-transmission rates, using RLNC for a priori and posteriori FEC, based on the quality of the connections in the network, and is shown to narrows down the trade-off between high-throughput and low-delay. This section is important since the it is the base of the communication solution that will be presented for the MS-MD interconnected networks.

This solution was first introduced in \cite{cohen2020adaptive}, and it is both casual, as it is reactive to the feedback acknowledgements (received after one round trip time, i.e., RTT, delay), and adaptive, as the rate of re-transmissions is adapted based on the varying channel conditions, and thus this scheme is called adaptive and casual RLNC. The error correction mechanism is a combination of both forward error correction (FEC) and feedback FEC (FB-FEC). The FEC mechanism sends re-transmissions to compensate for erasures that are predicted in advance to reduce the delay, while FB-BEC sends re-transmissions based on the received feedback to improve the throughput.
According to the actual link rates in the network, first, the source sends, a priori, an adaptive amount of FEC re-transmissions periodically. Then, at each transmission, according to posteriori re-transmission criterion, the source adaptively and causally decides over which paths to send FB-FEC re-transmissions and over which paths to send RLNC coded packets that contain new data information. The proposed re-transmission criterion is tracking the actual network packet receiving rate and the missing Degree of Freedom (DoF) rate required at the destination to decode the linear combination packets received as reflected from the feedback information.

We assume the forward channel from source to destination is erroneous, and the feedback channel is error-free. We aim at sending the coded packets such that the information packets can be decoded in-order and error-free at the destination while targeting a desired throughput and delay trade-off. In this section, we briefly review the AC-RLNC solution that was proposed for three different channel models in \cite{cohen2020adaptive,cohen2020adaptiveMPMH}: single P2P (SP), multipath (MP), and multipath multi-hop (MP-MH) communication channels. The error-free feedback and the fixed RTT assumptions are considered for simplicity, the model can be extended for the case where there are errors in the feedback channel and RTT fluctuations  e.g., by using the techniques provided in \cite{malak2019tiny}.

\subsubsection{Single Point-to-Point Communication Channel (SP)}\label{sec_background_ac_rlnc_sp}

Let consider a stream of information packets, i.e., $\{p_1,p_2\dots\}$. At each time step $t$, RLNC is performed over a contiguous window, called \textit{sliding window}, in the set of information packets, i.e, $c(t)=\sum_{i=w_\text{min}}^{w_\text{min}+w}\mu_ip_i$. The information packets involve in the linear coded packets are selected in sliding window mechanism to reduce delay and not in a large block of information as in traditional coding schemes, \cite{ho2006random}. Moreover, at the AC-RLNC solution defined a maximum window size for the sliding window, denoted by $\overline{o}$, to limit the maximum new packets of information that may involve in the linear coded combinations, i.e., $w\leq\overline{o}$. This limit mechanism is used to bound the maximum delay. If the value of $w_\text{max}=w_\text{min}+w$ at time slots $t$ and ($t-1$) does not change, the packet $c(t)$ is called a re-transmission and denoted by REP-RLNC packet, otherwise, it is called a new-transmission and denoted by NEW-RLNC packet. In other words, $w_\text{max}$ is incremented by each new-transmission, and $w_\text{min}$ is incremented once the sender ensures that the decoder was able to decode each $p_i$, i.e., $i\leq w_\text{min}$ (so-called in-order delivery).

\begin{figure}
    \centering
    \begin{tikzpicture}[align=center,node distance=7cm,>=stealth',bend angle=45,auto]

\tikzstyle{node}=[circle,thick,draw=gray!75,fill=gray!20,minimum size=2cm]

\tikzset{every loop/.style={min distance=3.3cm,in=-10,out=10,looseness=10}}

\begin{scope}
    \node [node] (src){Sender};
    \node [node] (dst) [right of=src]{Receiver};
    \draw [->,thick] (src) to [out=25,in=155] (dst);
    \node[draw,rectangle,thick,draw=gray!75,fill=white, minimum size=0.5cm] at (3.5,1.1){$\epsilon$};
    \draw [->] (dst) to [out=205,in=335] (src);
    \draw [dashed,thick,->] (src) to[loop right,in=-4,out=4,min distance=5.3cm,looseness=50] node[auto] {RTT} (src);
    \node[draw,rectangle,thick,draw=white,fill=white, minimum size=0.5cm] at (3.5,-1.0){ACK/NACK};
    \node [node] (pkt) [left of=src,node distance=5cm,draw=white,fill=white,minimum size=1cm]{\large$c(t)=\sum_{i=w_\text{min}}^{w_\text{min}+w}\mu_ip_i$};
    \draw [->] (pkt) to [out=0,in=180] (src);
\end{scope}

\end{tikzpicture}\vspace{-1cm}
    \caption{System model and AC-RLNC encoding process for an SP communication channel.}
    \label{fig:SP_AC_RLNC}
\end{figure}
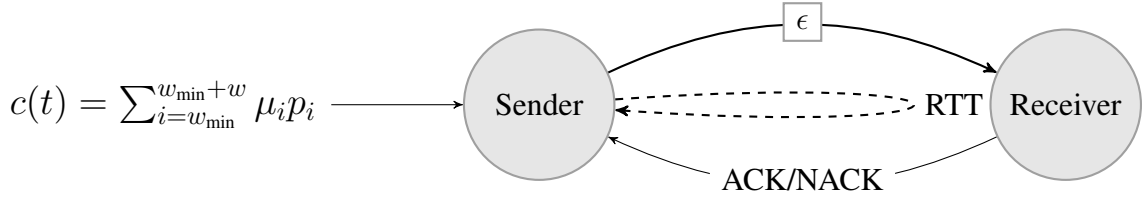

Fig.~\ref{fig:SP_AC_RLNC} shows the system model and encoding idea of AC-RLNC in an SP network. The forward channel is considered a BEC($\epsilon$), where $\epsilon$ may change over the time, and the backward channel transmits error-free ACKs/NACKs that are received at the sender at time slot $t$ for the packet that was transmitted at time $t-$RTT. The sender tracks the the channel state and decides whether to send a new-transmission or a re-transmission at each time step.

\subsubsection{Multipath Communication Channel (MP)}\label{sec_background_ac_rlnc_mp}

For the MP network, a new \emph{adaptive discrete water-filling algorithm} is proposed at the source nodes for the allocation of the new coded packets of information and the re-transmissions over the available global paths \cite{cohen2020adaptiveMPMH}. This adaptive discrete water-filling algorithm balances between two  objectives, maximizing throughput and minimizing the in-order delay, deciding the allocation of new and re-transmission packets to obtain the desired throughput-delay trade-off Fig. \ref{fig:Classical_vs_RLNC3}, bottom panel, illustrates the adaptive allocation scheme at the source nodes, when blue and green packets denote new and re-transmission packets of linear combination in the sliding window, respectively.

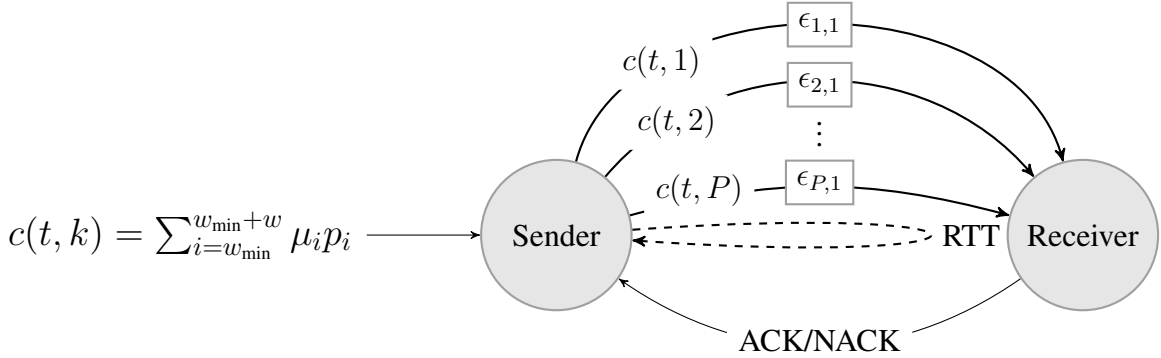
\begin{figure}
    \centering
    \begin{tikzpicture}[align=center,node distance=7cm,>=stealth',bend angle=45,auto]

\tikzstyle{node}=[circle,thick,draw=gray!75,fill=gray!20,minimum size=2cm]

\tikzset{every loop/.style={min distance=3.3cm,in=-10,out=10,looseness=10}}

\begin{scope}
    \node [node] (src){Sender};
    \node [node] (dst) [right of=src]{Receiver};
    \draw [thick,->] (src) to [out=75,in=105] (dst);
    \node[draw,rectangle,thick,draw=gray!75,fill=white, minimum size=0.5cm] at (3.5,2.8){$\epsilon_{1,1}$};
    \node[draw,rectangle,thick,draw=white,fill=white, minimum size=0.5cm] at (1.4,2.3){$c(t,1)$};
    \node[draw,rectangle,thick,draw=white,fill=white, minimum size=0.5cm] at (3.5,1.45){$\vdots$};
    \draw [thick,->] (src) to [out=50,in=130] (dst);
    \node[draw,rectangle,thick,draw=gray!75,fill=white, minimum size=0.5cm] at (3.5,2.0){$\epsilon_{2,1}$};
    \node[draw,rectangle,thick,draw=white,fill=white, minimum size=0.5cm] at (1.6,1.5){$c(t,2)$};
    \draw [thick,->] (src) to [out=15,in=165] (dst);
    \node[draw,rectangle,thick,draw=gray!75,fill=white, minimum size=0.5cm] at (3.5,0.7){$\epsilon_{P,1}$};
    \node[draw,rectangle,thick,draw=white,fill=white, minimum size=0.5cm] at (1.9,0.6){$c(t,P)$};
    \draw [->] (dst) to [out=215,in=325] (src);
    \draw [dashed,thick,->] (src) to[loop right,in=-4,out=4,min distance=5.3cm,looseness=50] node[auto] {RTT} (src);
    \node [node] (pkt) [left of=src,node distance=5cm,draw=white,fill=white,minimum size=1cm]{\large$c(t,k)=\sum_{i=w_\text{min}}^{w_\text{min}+w}\mu_i p_i$};
    \draw [->] (pkt) to [out=0,in=180] (src);
    \node[draw,rectangle,thick,draw=white,fill=white, minimum size=0.5cm] at (3.5,-1.4){ACK/NACK};
\end{scope}

\end{tikzpicture}\vspace{-0.7cm}
    \caption{System model and AC-RLNC encoding process for an MP communication channel.}
    \label{fig:MP_AC_RLNC}
\end{figure}

Fig.~\ref{fig:MP_AC_RLNC} shows the system model and encoding idea of AC-RLNC in an MP network. The forward channel is considered as composition of $P$ parallel independent BEC channels with parameters $\{\epsilon_{1,1},\dots,\epsilon_{P,1}\}$ (varying over time), and the backward channel transmits error-free ACKs/NACKs that are received at the sender at time slot $t$ for the packets that were transmitted at time $t-$RTT. The sender tracks the channel state and determines a subset of paths for re-transmissions, and the rest of paths for new-transmissions. The process of assigning a new-transmission or a re-transmission to each path is done via a modified water-filling algorithm that takes into accounts both the throughput and in-order delivery delay targets in optimization.

\subsubsection{Multipath Multi-Hop Communication Channel (MP-MH)}\label{sec_background_ac_rlnc_mpmh}

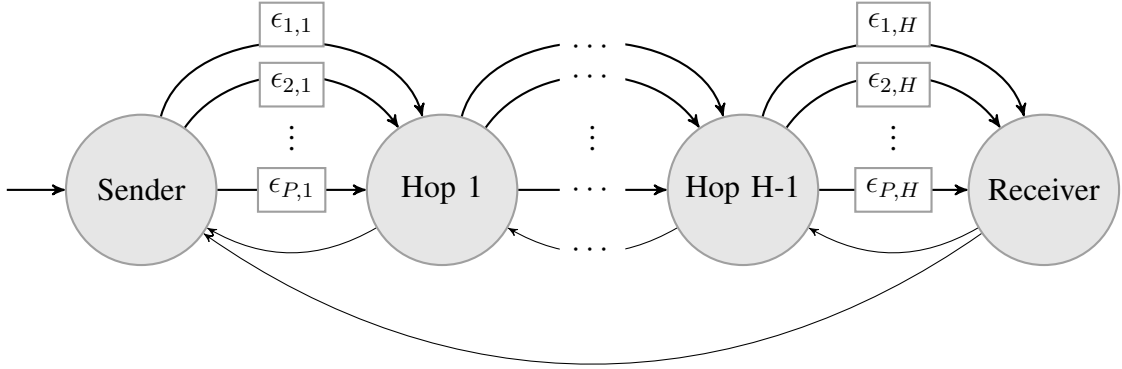
\begin{figure}
    \centering
    \begin{tikzpicture}[align=center,node distance=4cm,>=stealth',bend angle=45,auto]

\tikzstyle{node}=[circle,thick,draw=gray!75,fill=gray!20,minimum size=2cm]

\tikzset{every loop/.style={min distance=3.3cm,in=-10,out=10,looseness=10}}

\begin{scope}
    \node [node] (src){Sender};
    \node [node] (H1) [right of=src]{Hop 1};
    \node [node] (Hlast) [right of=H1]{Hop H-1};
    \node [node] (dst) [right of=Hlast]{Receiver};
    \draw [thick,->] (src) to [out=75,in=105] (H1);
    \node[draw,rectangle,thick,draw=gray!75,fill=white, minimum size=0.5cm] at (2,2.2){$\epsilon_{1,1}$};
    \node[draw,rectangle,thick,draw=white,fill=white, minimum size=0.5cm] at (2,0.8){$\vdots$};
    \draw [thick,->] (src) to [out=55,in=125] (H1);
    \node[draw,rectangle,thick,draw=gray!75,fill=white, minimum size=0.5cm] at (2,1.4){$\epsilon_{2,1}$};
    \draw [thick,->] (src) to [out=0,in=180] (H1);
    \node[draw,rectangle,thick,draw=gray!75,fill=white, minimum size=0.5cm] at (2,0.0){$\epsilon_{P,1}$};
    \draw [->] (H1) to [out=215,in=325,bend left=30] (src);
    
    \draw [thick,->] (H1) to [out=75,in=105] (Hlast);
    \node[draw,rectangle,thick,draw=white,fill=white, minimum size=0.5cm] at (6,1.9){$\cdots$};
    \node[draw,rectangle,thick,draw=white,fill=white, minimum size=0.5cm] at (6,0.8){$\vdots$};
    \draw [thick,->] (H1) to [out=55,in=125] (Hlast);
    \node[draw,rectangle,thick,draw=white,fill=white, minimum size=0.5cm] at (6,1.5){$\cdots$};
    \draw [thick,->] (H1) to [out=0,in=180] (Hlast);
    \node[draw,rectangle,thick,draw=white,fill=white, minimum size=0.5cm] at (6,0.0){$\cdots$};
    \draw [->] (Hlast) to [out=215,in=325,bend left=30] (H1);
    \node[draw,rectangle,thick,draw=white,fill=white, minimum size=0.5cm] at (6,-0.8){$\cdots$};  
    
    \draw [thick,->] (Hlast) to [out=75,in=105] (dst);
    \node[draw,rectangle,thick,draw=gray!75,fill=white, minimum size=0.5cm] at (10,2.2){$\epsilon_{1,H}$};
    \node[draw,rectangle,thick,draw=white,fill=white, minimum size=0.5cm] at (10,0.8){$\vdots$};
    \draw [thick,->] (Hlast) to [out=55,in=125] (dst);
    \node[draw,rectangle,thick,draw=gray!75,fill=white, minimum size=0.5cm] at (10,1.4){$\epsilon_{2,H}$};
    \draw [thick,->] (Hlast) to [out=0,in=180] (dst);
    \node[draw,rectangle,thick,draw=gray!75,fill=white, minimum size=0.5cm] at (10,0.0){$\epsilon_{{\small P,H}}$};
    \draw [->] (dst) to [out=215,in=325,bend left=30] (Hlast);
    
    \draw [->] (dst) to [out=225,in=315,bend left=35] (src);
    
    \node [node] (pkt) [left of=src,node distance=2cm,draw=white,fill=white,minimum size=0.2cm]{};
    \draw [thick,->] (pkt) to [out=0,in=180] (src);
\end{scope}

\end{tikzpicture}\vspace{-0.7cm}
    \caption{System model for an MP-MH communication channel.}
    \label{fig:MPMH_AC_RLNC}
\end{figure}

The MP-MH channel is more general than MP channel and is depicted in Fig.~\ref{fig:MPMH_AC_RLNC}. Each forward link can be considered as an independent BEC with parameter $\epsilon_{p,h}$ (varying over the time), where $p\in\{1,\dots,P\}$ and $h\in\{1,\dots,H\}$. The parameters $P$ and $(H+1)$ are the number of paths between two consecutive hops and the number of hops (including the sender and the receiver), respectively. In Fig~\ref{fig:MPMH_AC_RLNC}, it is assumed there are exactly $P$ forward links between any two consecutive hops. In this document, we extend the model to the case where the number of forward links between two consecutive hops are not fixed. There is an error-free local feedback link between two consecutive hops, and an error-free global feedback link from the receiver to the sender. Each hop can estimate the erasure probabilities of its incoming and outgoing links based on the received forward messages and local feedback messages, respectively. In the AC-RLNC solution presented in \cite{cohen2020adaptiveMPMH}, $P$ global paths are determined from the sender to the receiver, and then the the AC-RLNC solution presented in Section~\ref{sec_background_ac_rlnc_mp} is deployed over these $P$ global paths to efficiently transmit NEW-RLNC packets over a subset of global paths and send REP-RLNC packets over the rest.

In the MP-MH networks, the rates in the naive connections between the source and the destination are limited by the link with the lowest rate. This causes a bottleneck if the rate of the constituent links of a path have dramatic variations. Hence, for the MP-MH networks in \cite{cohen2020adaptiveMPMH}, is proposed a decentralized \emph{balancing algorithm} that avoids the throughput degradation that may be caused by the bottleneck effects.
This is obtained by reorganizing the selection of the naive global paths between the source and the destination to a new decentralized \emph{global path} selection optimization at the intermediate nodes (so-called ReEnc nodes), considering the rates of the incoming and outgoing links at each ReEnc node, to reduce the bottleneck effects.

The process of randomly mixing the incoming RLNC packets to prepare another set of RLNC packets is called \textit{traditional mixing}. By nature, all re-encoded packets generated in an intermediate node through traditional mixing are NEW-RLNC packets even if there was only one NEW-RLNC packet exist among the incoming packets to the node. Traditional mixing results in an increased in-order delivery delay. To increase the efficiency, in terms of maximizing throughput and minimizing in-order delivery delay, the ReEnc nodes perform a \emph{selective mixing} in which the new-transmissions at each time step are re-encoded together and the received re-transmissions are also re-encoded together. In the bottom panel of Fig. \ref{fig:Classical_vs_RLNC3}), the links transmitting new-transmissions and re-transmissions using the selective mixing are marked in blue and green, respectively. This balancing algorithm together with the selective mixing solution closes the mean and max in-order delay gap and boosts the throughput. Thus, AC-RLNC for MP-MH network with one source and one destination can reach the desired delay-throughput trade-off.

For the non-asymptotic regime, AC-RLNC solution for MP-MH networks as proved in \cite{cohen2020adaptiveMPMH} may achieve more than $90\%$ of the network capacity with zero error probability under mean and maximum in-order delay constraints. Fig.~\ref{fig:Classical_vs_RLNC3} demonstrates a comparison between TCP, traditional RLNC, and AC-RLNC communication communication solutions over MP-MH networs. We remark that in this paper, we also investigate the case of an MP-MH communication channel where a subset the intermediate nodes are not capable of performing AC-RLNC solution such as performing the  re-encoding and balancing. This is an important feature since it makes the possibility of having a highly-heterogeneous network where a subset of nodes are not yet capable of adapting to the new technology, and consequently, provides a smooth transition phase. Incorporating the traditional nodes necessitate some  modifications and considerations in the AC-RLNC solution that will be discussed in detail in Section~\ref{sec_architecture}.

The AC-RLNC solution proposed in \cite{cohen2020adaptive} and  \cite{cohen2020adaptiveMPMH} can reach the desired throughput-delay trade-off for an MP-MH network with one source and one destination, and showed to be very efficient in practice. However, modern communications have moved away from point-to-point models to increasingly heterogeneous highly-meshed networks with multiple sources and destinations that share the interconnected communication medium. Hence, to address the demands in the modern communications, in Sections \ref{sec_main_controller} and \ref{sec_modules_cooperation}, we propose a novel controller-based protocol to demonstrate how it possible to deploy AC-RLNC solution in heterogeneous highly-meshed communication networks.

\begin{figure}
    \centering
    \includegraphics[trim=0cm 0.0cm 0cm 0cm,width = 1 \columnwidth]{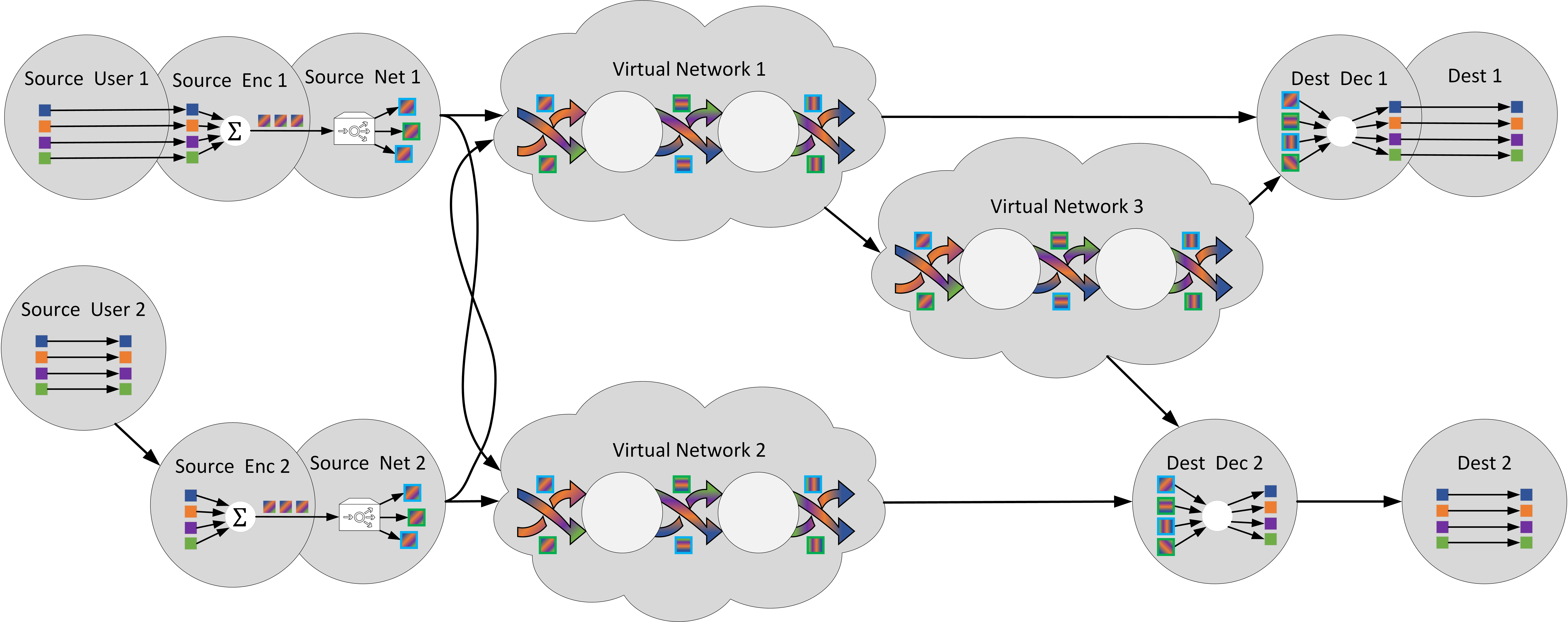}
    \caption{A simple network architecture that is composed of three virtual networks, and has two sources and two destinations. \off{\textcolor{blue}{We may use another terminology for 'User' to be more consistent with 'Dest' node and also to consider both uplink and downlink scenarios.}}}
    \label{fig_MP_MH_in_Bigger_Net}
\end{figure}

\section{Network Architecture for a Heterogeneous MS-MD communication}\label{sec_architecture}

In this section, we present our new architecture for an MS-MD network of interconnected nodes. Each source intends to send messages to a subset of destinations through the shared network. We divide the network into several \textit{Virtual Networks (VNs)}, where node arrangements are managed by the SSE-SDNC. Each VN has one communication channel to transfer the received packets, and the communication channel can be MP-MH (in the general case). We describe the main parts of the proposed network architecture through the example depicted in  Fig.~\ref{fig_MP_MH_in_Bigger_Net} , with two sources, two destinations, and three VNs. In this example, Source~1 can transmit information to both destinations, however, Source~2 can only transmit information to Destination~2. Each source is associated with one \textit{User}, one \textit{Enc node}, and one \textit{Net node}, and each destination is associated with one \textit{Dec node} and one \textit{Dest node}. Each source is aware of the available paths and routing information for the current application to its desired destinations through the interaction with messaging systems, where the SSE-SDNC provides the updated values of the information regarding paths and links characteristics.

User generates the application/service information, and it can be located at the same place as the source nodes that generate the RLNC coded data, see Source~User~1, or it can be physically apart, see Source~User~2. In latter, User may deploy traditional communication methods to transmit the information to the first place on paths to its desired destination that supports the AC-RLNC solution, i.e., Enc and Net nodes. An Enc node prepares the RLNC-coded packets based on the received feedbacks and the tracked channel status. A Net node selects the global paths over which the new-transmissions are performed, called global paths of type-1, and subsequently, global paths over which the re-transmissions are transmitted, called global paths of type-2. Dec node is the closest to the destination that can perform decoding. The Dec node performs decoding, provides necessary feedback, and transmits the decoded packets to the destination. Finally, we use traditional communication protocol from Dec node to the destination. Dest node represent the physical location of the final destination.

\begin{figure}
    \centering
    \includegraphics[trim=0cm 0.0cm 0cm 0cm,width = 0.9 \columnwidth]{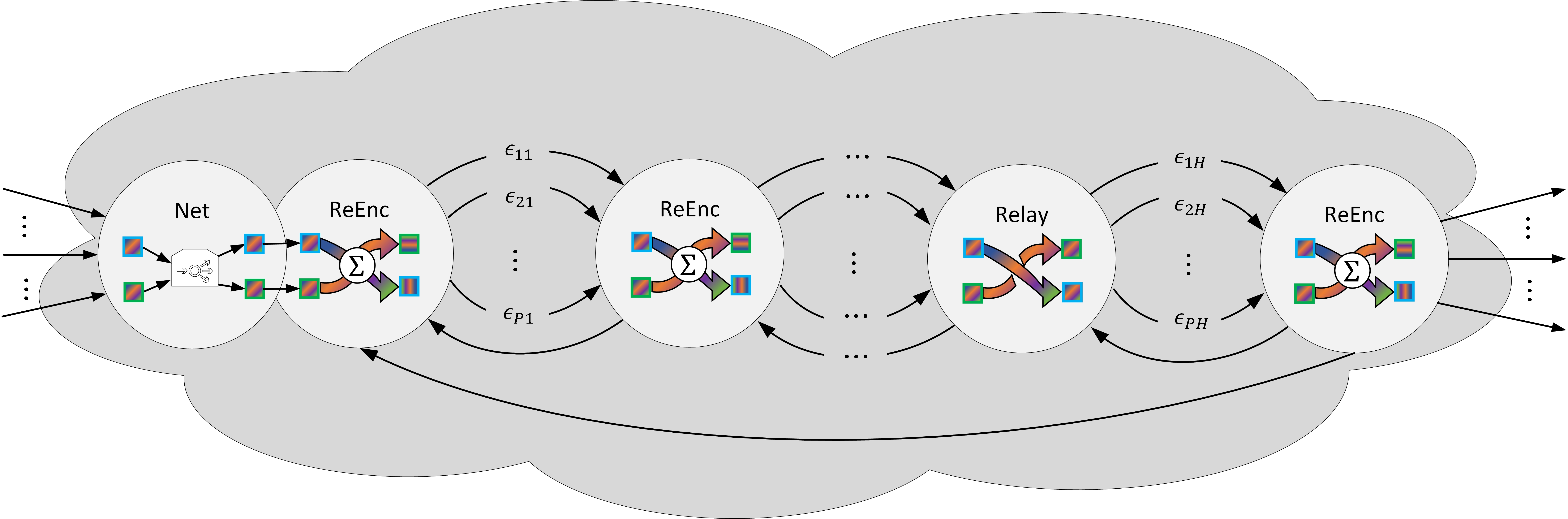}
    \caption{The general architecture of a virtual network. When $H=1$, the network topology we be reduced to an MP network. When $H=1$ and $P=1$, the network topology we be reduced to an SP network.}
    \label{fig_MP_MH}
\end{figure}

We next introduce the main parts of a VN and their tasks. Fig.~\ref{fig_MP_MH} depicts the system model for a VN in the general case of MP-MH topology. In a VN,  \textit{Net node} is responsible to select global-paths of type~1 and type~2. Then, other nodes are either a \textit{ReEnc node} or a \textit{Relay node}. A ReEnc node performs selective mixing (if supported) over the received RLNC packets and transmits the re-encoded RLNC packets on appropriate global-paths, or it performs traditional mixing (if supported) and transmits the re-encoded RLNC packets on global paths (all with same type). The Relay Nodes forward the received packets on their global paths whether they are re-transmissions or not using routing table.

\section{Scalable, Secure and Efficient SDN Controller for Meshed MS-MD Communication}\label{sec:Controller}

\begin{figure}
    \centering
    \includegraphics[trim=0cm 0.0cm 0cm 0cm,width = 1 \columnwidth]{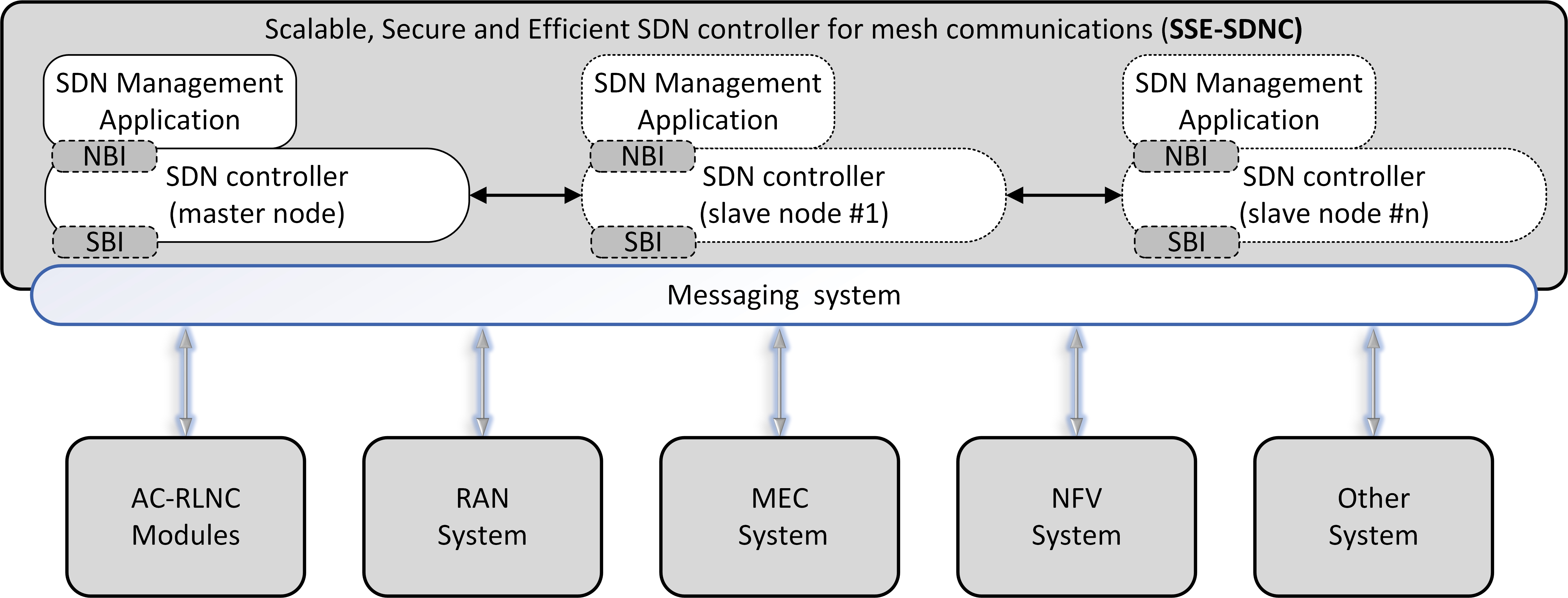}
    \caption{Scalable, Secure and Efficient SDN controller for mesh communications (SSE-SDNC) and messaging system.}
    \label{fig_controller_mess_sys}
\end{figure}

SDNs characterize network architecture by separating control plane from data plane and providing support for heterogeneous network interplay with rapid evolution and dynamism using programmable planes. Keeping up with rapid evolution and dynamism of network traffic SDN is a crucial element of 5G networks and beyond, along with other technologies like Network Function Virtualization (NFV). A SDN controller provides flexible management and control of network devices, thus it has the necessary information for the AC-RLNC solution over heterogeneous meshed networks with multiple sources and destinations. The Scalable, Secure and Efficient SDN Controller (SSE-SDNC) for mesh communications relies on multiple SDN controllers to attain an improved fault tolerance. The constituent controllers are organized in a cluster, so that the master node can be easily replaced by other slave nodes, as depicted in Fig.~\ref{fig_controller_mess_sys}. The cluster of SDN controllers can be managed using available solutions like the Atomix ONOS cluster~\cite{Xie2019}. Within such approach the information regarding the network topology is distributed and accessible by the diverse controllers.

The SSE-SDNC, using multiple SDN controllers, manages the topology information of the virtual networks through the southbound interface (SBI), for instance, by supporting OpenFlow, P4, or NETCONF protocols. Such protocols allow to control the forwarding plane (e.g. flow rules in devices) and to manage devices (e.g. software configuration and updates).
The SSE-SDNC also includes management application components that communicate with the SDN controller through the northbound API (NBI), supporting REST or gRPC technologies. The management application is responsible to retrieve the information from SDN controllers, regarding the network topology (e.g. RTT) and provide it to other components like the AC-RLNC modules.
The information exchange between the SSE-SDNC and other components is performed through the messaging system, using YANG data models. The following subsections detail some of the solutions for each interface.

\subsection{SDN controllers for mesh networks in 5G}
The architecture of a SDN controller includes an application layer, responsible to interact with management applications (see section~\ref{subsec:SDNController:NBI}), infrastructure layers, which is responsible to control and manage devices (see section~\ref{subsec:SDNController:SBI}). A central layer includes the SDN controller with  the management functionalities, and some controllers also include a Eastwestbound interface to allow the communication with other components or SDN controllers.

There are a plethora of SDN controllers available and the majority is released as open-source projects, as surveyed in~\cite{Mamushiane2018,Amin2018}. Some of these SDN controllers, worth mentioning include Ryu, OpenDayLight, ONOS and Floodlight. The main difference between them relies on the architecture (if running centralized, distributed or in a hierarchical fashion), the underlying programming language, the supported protocols in the distinct interfaces, the support for scalability, reliability and consistency. Given the requirements of the SSE-SDNC, Ryu and Floodlight are not suitable solutions due the centralized architecture model (i.e. introducing a single point of failure - SPOF), and do not scale to support large SDN deployments (e.g. networks in a smart city scenario) and have limited reliability support~\cite{Bannour2018}. On the hand, OpenDaylight and ONOS are able to support distributed deployments and large SDN deployments, and have reliability support as required in the smart city scenario. In addition, the performance of these controllers does not diverge too much~\cite{Zhu2019,Badotra2019}. For instance, OpenDayLight is able to provide a better performance regarding time to process flows, while ONOS is able to support higher flow response rate (flows/ms).

In a security perspective, either ONOS and OpenDayLight controllers have Denial of Service (DoS) and authentication vulnerabilities~\cite{Secci2019}, with OpenDayLight presenting a higher number of vulnerabilities.

\subsection{North Bound Interface - NBI}
\label{subsec:SDNController:NBI}
The North Bound Interface (NBI) of SDN controllers allow the bidirectional communication between controllers and management applications. Either for retrieving information regarding the status of network topology, as well as to configure the behaviour of the SDN controller when managing the flows tables in the networking devices (e.g. switches, routers).

The NBI commonly relies on REST interface, with the unidirectionality drawback. ONOS includes support for gRPC for bidirectional communications, which can be an advantage for SDN management applications, as those planned in the SSE-SDNC. Through the NBI, the SDN Management Application can configure the behaviour of the SDN controller, for instance by using available APIs, such as the intent or flow objective API, which allow to install flow rules in the devices by only requiring application to express their objective or intent (e.g. that a source node is able to communicate with a destination node). By expressing such intents, the SDN controller is able to install/manage the necessary flow rules in network devices.

\subsection{South Bound Interface - SBI}
\label{subsec:SDNController:SBI}
The South Bound Interface (SBI) of SDN controllers is employed to manage, control networking devices such as switches, routers employing standard protocols like OpenFlow, P4 or NETCONF. The difference between OpenFlow and P4 is related with the fact that P4 is a programming language for the data plane in networking devices and Network Interface Cards (NICs) on end-nodes, as opposed to OpenFlow which allows the management of the forwarding and control data plane~\cite{Liu2018}.

P4 introduces flexibility in the configuration of the devices, since the behaviour of common functionalities can be programmed with P4 language. For instance the forwarding behaviour in switches can be implemented in a P4 program, without requiring the explicit management of flow tables through the control plane like in OpenFlow. In this regards, several projects like the H2020 5Growth are employing P4 to program the 5G infrastructure to support end-to-end services~\cite{5growth_d22}. Another advantage of employing P4 is related with the network telemetry, since the collection of information regarding networking devices can be included in the normal data packets, avoiding the necessity of using control messages to gather information~\cite{Guan2019}. Despite the increased flexibility, networking devices and NICs neet to run P4Runtime, which is responsible to run the P4 programs and to communicate with the SDN controllers~\cite{p4Lang}.

The control and management of networking devices is also promoted with specific layers, implementing switch operating system, like stratum~\cite{stratum} which facilitates the management of devices in a scalable fashion.

\subsection{Messaging System and Other Systems}
\label{subsec:SDNController:MessagingSystem}
The SSE-SDNC interacts with other systems and with the AC-RLNC modules through a messaging system, as illustrated in Fig.~\ref{fig_controller_mess_sys}. The messaging can rely on publish subscribe (e.g. MQTT) or data stream models (e.g. kafka), which support communication between different systems in a distributed and reliable fashion.

In this regard the SDN management applications interacting the distributed SDN controllers can provide information regarding the network, events (e.g. new device available) through the messaging system, as a Kafka stream ~\cite{onos_apps_kafka}. The advantage of relying in such kind of messaging is that available implementations of MQTT or Kafka provide fault tolerance by supporting multiple brokers to handle the diverse exchanged messages.

The information models exchanged through the messaging system will also comply with current standardization efforts~\cite{Janz2016}. The terminology and the diverse models specified in the ONF Common Information Model (ONF-CIM), in the ONF Core Network Model (ONF-CNM), among others will be employed to further enhance the interoperability between SDN solutions and other systems. Such information models also include support for YANG, JSON data representation. The transport API (TAPI)~\cite{ONF_TAPI} is also relevant to consider between the diverse systems.

\section{AC-RLNC in conjunction with controller for meshed MS-MD Communication}\label{sec:MS-MD-AC-RLNC}

In this section, we present an AC-RLNC solution for the heterogeneous MS-MD network architecture. The solution was previously introduced in \cite{cohen2020adaptive,cohen2020adaptiveMPMH}, and in this article, we propose a practical implementation for an MD-SC network, where here some necessary modularization, modification, and interplay with main controller are incorporated, and thus, resulting in a decentralized implementation. This is a necessary step toward practical utilization of network coding in meshed heterogeneous networks as each module requires specific information and needs to be implemented in a specific part of the meshed network. We remind that the presented AC-RLNC solution is an example of how a distributed communication/computation solution can be deployed over a highly-meshed MS-MD network, by presenting effective architecture and protocols. One can use a similar approach to deploy other solutions, e.g., TCP communication, over the presented MS-MD network architecture.

\subsection{Main Modules of AC-RLNC Solution for MS-MD networks}\label{MP_MH_in_Bigger_Net}

In this subsection, we describe the main modules of an AC-RLNC solution for the heterogeneous MS-MD network architecture, how they work together, and in which node in the proposed network architecture they need to be implemented. Then, we discuss the information each module requires to perform its tasks, and how this information is provided through interacting with the SSE-SDNC controller, communicating with other modules, and/or feedback. The main modules of the proposed AC-RLNC solution for heterogeneous MS-MD networks are the following:

\begin{itemize}
    \item \textbf{Agent (AGN):} This module is present at all network nodes that require interaction with the controller, and is responsible to retrieve necessary  information from SSE-SDNC, e.g., Fairness Table, RTT, Link Rates, through the messaging system, and/or to determine the information locally in the node like the Routing Table.
    \item \textbf{Balancing (BAL):} This modules matches the incoming links and outgoing links such that the bottleneck effect is minimized over each global path. The balancing module is implemented at ReEnc nodes.
    \item \textbf{Global Path Identification (GP):} The role of this module is identifying global paths between two parts of the network, and it is implemented at Net nodes in the proposed network architecture. In a Source Net node, the module identifies the available global paths for the current application/service from Source to Destination. In a VN Net node, it identifies the available global paths for the given application/service from the first to the last node in the VN.
    \item \textbf{Budgeting (BUG):} This module splits the rate budget (decides which global-paths are type~1 and which are type~2) for an application/service into re-transmissions and new-transmissions, and it is implemented at Net nodes.
    \item \textbf{Encoding (ENC):} This module performs RLNC encoding on the information packets of an application/service and prepares RLNC packets, and it is implemented at Source Enc nodes. We remind that each RLNC packet is either a re-transmission or a new-transmission.
    \item \textbf{Packet Allocation (PA):} The role of this module is allocating each global path of type~1 a new-transmission packet and each global path of type~2 a re-transmissin packet. This module is implemented at Source Net nodes and ReEnc nodes.
    \item \textbf{Re-encoding (REC):} This module linearly mixes the received RLNC packets and produces new RLNC packets, using selective mixing or traditional mixing. It is implemented at ReEnc nodes.
    \item
    \textbf{Decoding (DEC):} This module decodes the RLNC packets associated with the service/application that is desired by the destination and transmits necessary acknowledgment feedback messages. The decoding module is implemented at Dec node.
\end{itemize}

We describe the interaction between AC-RLNC modules and with the main controller in Section~\ref{sec_main_controller} and \ref{sec_modules_cooperation}, see also Fig.~\ref{fig_all_modules_scheme}. Fig. \ref{fig:networklayers} shows the OSI stack layers and nodes in which the modules are implemented. More details about the modules' implementations are given in Section~\ref{sec_modules_implementation}. We note that if all or a subset of AC-RLNC modules cannot be run in parts of the network nodes, the controller can perform the task of the missing modules and make necessary changes according to the results.

\subsection{Providing Necessary Information for Modules through SSE-SDNC}\label{sec_main_controller}

The SSE-SDNC is responsible for providing parts of necessary information for each module. Besides, the AGN module interacts with the SSE-SDNC to fetch the necessary information, and update necessary information. All information offered by the SSE-SDNC is not needed by every module in the presented solution, thus a set of distributed controllers, to have a low-delay module-controller communication, is also a promising idea - using the AGN modules, which communicates with the SSE-SDNC to synchronize updates in the network topology. The information that the AC-RLNC solution demands from the controller along with a brief explanation are listed below:


\subsubsection{Routing Table (RT)} Each RT is associated with a Net node, and it contains routing information (constituent links) between two IP addresses. In our solution, we need an RT per Net node. For example, an RT table that is requested by a Source Net node of an application/service contains all routing information from the source to the destination, however, an RT table that is requested by a VN Net node has all routing information from the first node to the last node in that VN. The AGN module is responsible for the RT information through interacting with the SSE-SDNC.

\subsubsection{Fairness Table (FT)} There exists one FT per Net node that shows how to split the available rate resources among several applications/services based on their priorities. For example, one can assign rate per application/service that is a real number in $[0,1]$ such that these real numbers sum up to $1$. One can also quantify the budget in several levels where the priority (urgency) of an application is quantified by a level. The SSE-SDNC is responsible for setting the weights/costs associated with the network resources. Then, AGN module presents this information to AC-RLNC modules at Net nodes.

\subsubsection{Local-Path Routing Table (LPRT)} There exists one LPRT per ReEnc node. An LPRT shows the matching between the incoming links and outgoing links, and it is constructed and updated by Balancing modules of a ReEnc node to minimize the bottleneck effect. The ReEnc nodes update their LPRTs based on dynamism of the network condition where the links quality may change over the time.  Traditional Relay nodes do not need to update their LPRTs, and this provides heterogeneity property of using traditional nodes along with the nodes that are capable of performing the AC-RLNC solution.
The AGN module is responsible for providing the RT information.

\subsubsection{Global-Path Routing Table (GPRT)}There is one GPRT per Net node of an application service. A GPRT includes all global paths routing information and rate between a pair of IP address, i.e., source and destination of an application/service when requested by a Source Net node, or first and last nodes of a VN when requested by a VN Net node. A GPRT is constructed and updated by Global path identification module with the assistance of the AGN module.

\subsubsection{Round Trip Time (RTT) and Link Rates} 
The SSE-SDNC is responsible for providing updated values of the RTTs and link rates, which are determined through the probing mechanisms implemented in the SDN management applications.

\begin{figure}
    \centering
    \includegraphics[trim=0cm 0.0cm 0cm 0cm,width = 1 \columnwidth]{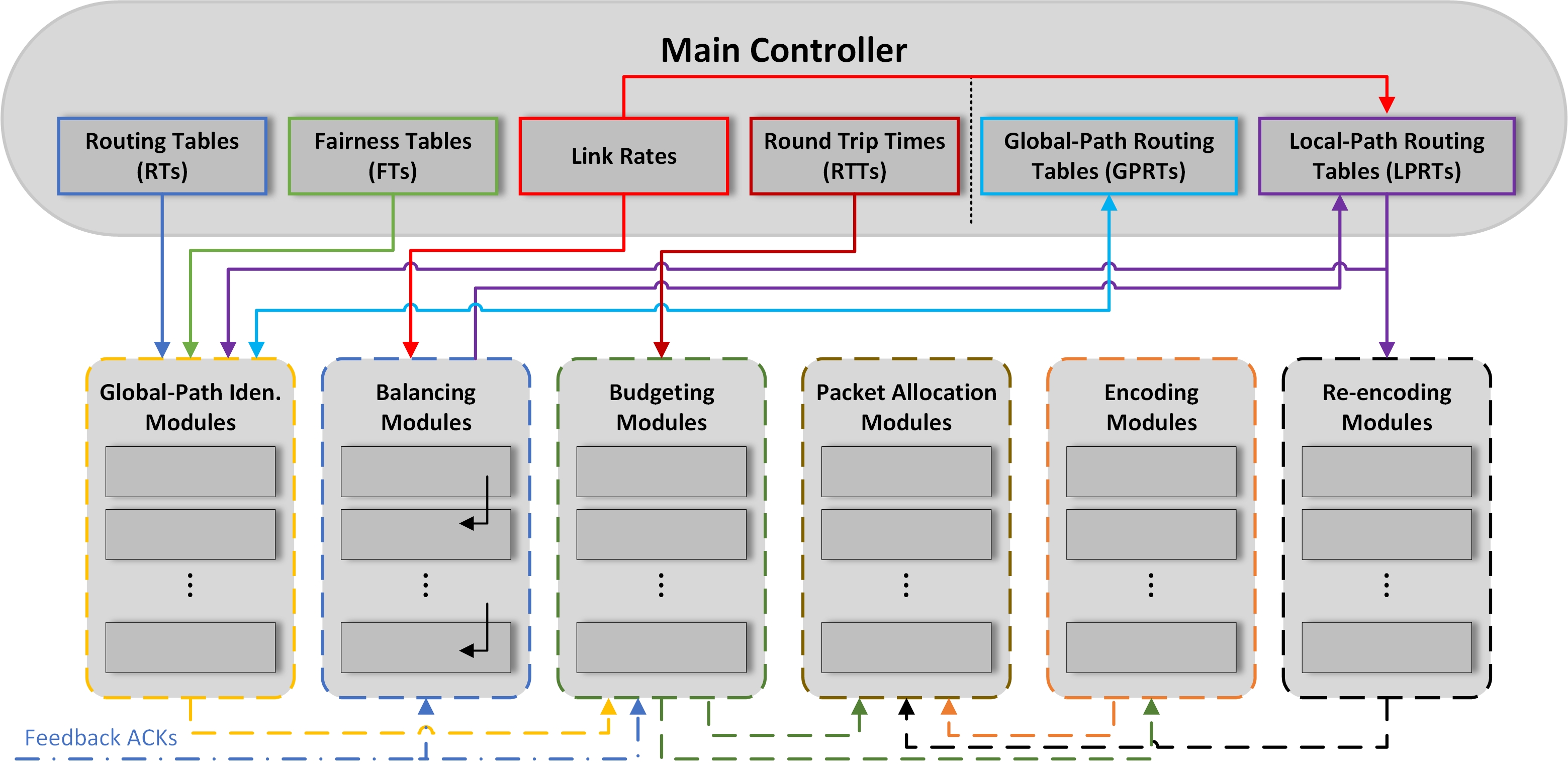}
    \caption{The interaction of AC-RLNC modules and the main controller to realize the AC-RLNC solution in a heterogeneous MS-MD network. Note that the modules of one type, e.g., balancing modules, are implemented in different parts of the network, e.g., individual ReEnc Nodes. Here, for sake of presentation, we group all modules of same type together. \off{\textcolor{blue}{[Color descriptions will be added]}}}
    \label{fig_all_modules_scheme}
\end{figure}

Here, we briefly describe information that each AC-RLNC module requires (from the SSE-SDNC, and/or AGN or other modules) to perform its tasks. More details and algorithmic implementations can be found in Section~\ref{sec_modules_implementation}. Fig.~\ref{fig_all_modules_scheme} illustrates a how different AC-RLNC modules and the controller exchange information during data transmission. The balancing module is called at a ReEnc node and requires the link rate for incoming and outgoing links (through the main controller, by probing, or from feedback), and construct/update the LPRT of the node at SSE-SDNC, see Algorithm~\ref{alg:Balancing}. For the global-path identification module, the required information depend on where it is called. At a Source Net node, the module requires RT, FT, and involved VN GPRTs to construct a new GPRT at the main controller. The GPRTs for the associated VNs (The VNs that carry information from Source to Destination) results in faster identification of global paths at a Source Net node. In a VN Net node, the module requires RT, FT, and LPRTs of constituent VN nodes, See Algorithm~\ref{alg:global_paths}. The budgeting module requires the GPRT (from the global path identification module) and RTT (from the main controller, feedback, or through a probing mechanism), see Algorithm~\ref{alg:Budgeting}. The encoding module requires the number of new-transmissions and re-transmissions along with the sliding window limits from the budgeting module, see Algorithm~\ref{alg:Encoding}. The re-encoding module performs re-encoding on the received RLNC packets based on the mixing mechanism (selective mixing, traditional mixing, or none) and prepares a new set of RLNC packets, see Algorithm~\ref{alg:Forward}. The decoding module performs decoding on received RLNC packets to retrieve the information packets when possible and also provides feedback. \off{\textcolor{blue}{[There is no separate algorithm for this module yet in Section~\ref{sec_modules_implementation}, and it is implemented as a part of Algorithm~\ref{alg:main_continue}]}.}

Algorithm~\ref{alg:main} shows how the SSE-SDNC initiates the AC-RLNC solution upon a new service/application request between a pair of User and Destination by collecting necessary information. Some initialization steps are performed by the SSE-SDNC (lines~1-15), and some are performed by AC-RLNC modules and the results are recorded in the SSE-SDNC (lines~16-31). After necessary initialisation steps, Algorithm~\ref{alg:main_continue} is called for the data transmission, line~32. Once the data transmission is done, necessary terminating steps are performed for the finished application/service request, lines~33-52.

\begin{algorithm}\small 
    \caption{Main AC-RLNC for SNOB 5G}
    \label{alg:main}
    \begin{algorithmic}[1] 
        \Statex Called when a user requests an application/service from the Main-Controller
        \Statex\textbf{Input:} User and Destination IP addresses, Prioritization level of requested service/application (if supported).  
  
       \Statex\textbf{\underline{Initialization at the Main-Controller}:}
        \State \textbf{Stage 1:} Routing Table (RT) 
        \If{no RT for (User, Destination)}
            \State Build an RT for (User, Destination).
        \EndIf
        \State Assign the RT to the new application/service.
        \State Specify the nodes involved in the new application/service and their roles given the RT.
        
        \State \textbf{Stage 2:} Fairness Table (FT)
        \ForEach{Net node involved with the new application/service}
            \If{no FT assigned to the Net node}
                \State Construct an FT with the new application/service and assign it to the Net node.
            \Else
                \State Add the new application/service to the FT.
            \EndIf
            
            \State Split the rate resources among the remaining applications/services (based on their priorities if supported).
        \EndFor
    \State \textbf{Stage 3:} Provide the relevant information to AC-RLNC modules (see  Fig.~\ref{fig_all_modules_scheme})
    \ForEach{VN Net involved with the new application/service}
        \ForEach{application/service in the FT}
            \State Identify the global paths and their rates using Algorithm \ref{alg:global_paths}.
        \EndFor
    \EndFor 
    
    \ForEach{VN Net involved with the new application/service}
        \ForEach{application/service in the FT}
            \State Match the incoming and outgoing links the nodes in the VN using Algorithm \ref{alg:Balancing}.
        \EndFor
    \EndFor
    
    \ForEach{Source Net node involved with the new application/service}
        \ForEach{application/service in the FT}
            \State Identify the global paths and their rates using Algorithm \ref{alg:global_paths}.
        \EndFor
    \EndFor 
    
    \Statex\textbf{\underline{Transmission of the data}}
    \State The data transmission algorithm is given in Algorithm  \ref{alg:main_continue}.
    
    \Statex\textbf{\underline{Termination at the Main-Controller}:}
    \ForEach{Net node involved with the finished application/service}
        \State Remove the finished application/service from FT and .
        \State Split the rate resources among the remaining applications/services (based on their priorities if supported).
    \EndFor
    \State Remove associated GPRTs and LPRTs from the main controller.
    \ForEach{VN Net involved with the finished application/service}
        \ForEach{application/service in the FT}
            \State Identify the global paths and their rates using Algorithm \ref{alg:global_paths}.
        \EndFor
    \EndFor 
    \ForEach{VN Net involved with the finished application/service}
        \ForEach{application/service in the FT}
            \State Match the incoming and outgoing links the nodes in the VN using Algorithm \ref{alg:Balancing}.
        \EndFor
    \EndFor
    
    \ForEach{Source Net node involved with the finished application/service} 
        \ForEach{application/service in the FT}
            \State Identify the global paths and their rates using Algorithm \ref{alg:global_paths}.
        \EndFor
    \EndFor 
   \end{algorithmic}
\end{algorithm}

The main controller performs the necessary initialisation in three stages. At first stage, the main controller checks if there already exists a routing table (RT) between User and Destination IP addresses, and if not, it constructs a new RT. Then, the RT is assigned to the new application/service (lines~1-5). Next, the main controller identifies the nodes that perform the AC-RLNC solution, i.e., Source Enc and Net nodes, VN Net nodes, ReEnc nodes, Relay nodes, and Dest Dec node involved in the current service/application using the RT (line 6). We remind that each application/service request needs one Enc node, one Source Net node, one or more VNs, and one Dec node. Per VN that carries information for the application/service, there are one Net node, and a number of Relay nodes and ReEnc nodes.

At second stage, the main controller adds the new application/service to the fairness table (FT) associated with each Net node that is involved with the new application/service, i.e., Net nodes that are on the routing paths of the new application/service. For the involved Net nodes that do not already have a FT, one with the new application/service is constructed. Then, the main controller splits the rate resources among the applications/services in each FT by updating priorities (lines~7-15). We note that a source can be assigned to more than one applications/services, and similarly a VN may carry information for more than one sources (consequently application/services). How to split the available rate resources (global paths) among the applications/services is determined in Net nodes using the FT. If the application/service prioritization is not supported in a Net node, the rate resources are split equally among the applications/services in the FT, including the new service/application. Otherwise, a discrete water-filling optimization can be deployed to distribute the global paths among the services/applications.

At third stage, the main controller calls the balancing modules and global-Path Identification modules to construct/update related GPRTs and LPRTs, lines 16-31. At this stage, one GPRT is constructed for the new application/service per Net node, and the remaining GPRTs associated with the involved Net nodes are updated accordingly. Similarly, one LPRT is constructed for the new application/service per VN node, and the remaining LPRTs associated with the involved VN nodes are updated accordingly. The main controller provides necessary information for the modules. This includes providing related RTs and FTs to global path identification modules, link rates to balancing modules, and RTTs to budgeting modules. First, for each involved VN Net node, the global path identification module is called for all applications/services in the FT (lines~17-21). Next, for each involved VN Net node, the balancing module is called for all applications/services in the FT (lines~22-26). Then, for each involved Source Net node, the global path identification module is called for all applications/services in the FT (lines~27-31).

Once the main controller finishes the initialization, Algorithm~\ref{alg:main_continue} is called to perform the application/service data transmission using the AC-RLNC solution (line 32), which will be describe in detail in Section~\ref{sec_modules_cooperation}.

Once the application/service data transmission ends, the main controller performs necessary termination steps as follows. First, the main controller removes the finished application/service from the fairness table (FT) associated with each Net node that is involved with the finished application/service. Then, the main controller splits the rate resources among the the remaining applications/services in each FT by updating priorities (lines 33-36). Next, the main controller removes the GPRTs (resp., LPRTs) associated with the involved Net nodes (resp., VN nodes) for the finished application/service (line~37). Finally, the main controller calls the balancing modules and global-Path Identification modules to update related GPRTs and LPRTs. First, for each involved VN Net node, the global path identification module is called for all remaining applications/services in the FT (lines~38-42). Next, for each involved VN Net node, the balancing module is called for all applications/services in the FT (lines~43-47). Then, for each involved Source Net node, the global path identification module is called for all applications/services in the FT (lines~48-52).

\subsection{Cooperation of AR-RLNC modules to perform the complete solution}\label{sec_modules_cooperation}

In this subsection, we precisely describe how the AC-RLNC modules collaborate and also interact with each other and with SSE-SDNC to perform the error-free data transmission task over a heterogeneous MS-MD network.  Algorithm~\ref{alg:main_continue} shows how AC-RLNC modules perform AC-RLNC data transmission for an application/service altogether in a collaborative and distributed fashion. The operations performed at each node in the presented network architecture are described in more detail below:

\begin{algorithm}\small 
  \caption{Main AC-RLNC for SNOB 5G - Data Transmission}
  \label{alg:main_continue}
  \begin{algorithmic}[1] 
    \Statex Algorithm is called by Main AC-RLNC Algorithm
    \While{information packets for the application/service to transmit}
        \Statex\textbf{\underline{User}}
         \If{User and Source nodes (Source Enc and Net nodes) do not the have same IP address}
            \State Encode data in traditional IP packets and transmit them to Source Enc node using the RT. 
        \EndIf
    
        \Statex\textbf{\underline{Source Enc and Net nodes}}
        \State \textbf{Stage 1:} Specify the global paths of type~1 (for new-transmissions) and the global paths of type~2 (for re-transmissions) using budgeting module given in Algorithm~\ref{alg:Budgeting}.
        \State \textbf{Stage 2:} Prepare NEW-RLNC packets and REP-RLNC packets using encoding module given in Algorithm~\ref{alg:Encoding} 
        \State \textbf{Stage 3:} Assign the RLNC packets through appropriate global paths using packet allocation module given in Algorithm~\ref{alg:PacketAllocation}. 
        \Statex\textbf{\underline{VN Net Node}}
        \State Specify the global paths of type~1 (for new-transmissions) and the global paths of type~2 (for re-transmissions) using budgeting module given in Algorithm~\ref{alg:Budgeting}.
        \Statex\textbf{\underline{ReEnc and Relay Nodes}}
            \ForEach{VN Relay node}
                \State Forward coded packets using the RT.
            \EndFor
            \ForEach{VN ReEnc node}
                \State \textbf{Stage 1:} Prepare NEW-RLNC and REP-RLNC packets using re-encoding module given in Algorithm \ref{alg:Forward}.
                \State
            \textbf{Stage 2:} Assign the RLNC packets through appropriate global paths using packet allocation module given in Algorithm~\ref{alg:PacketAllocation}.
            \EndFor
        \Statex\textbf{\underline{Dec Node}}

        \State \textbf{Stage 1:} Decoding 
        \If{there are sufficient RLNC packets for decoding}
            \State Decode the RLNC packets \cite{pedersen2011kodo}.
            \If{Dec node and Destination do not have the same IP address}
                \State Embed the decoded data into traditional IP packets and transmit them to Destination using the RT. 
            \EndIf
        \Else
            \State Update counter as number of RLNC packets received toward decoding the first information packet that is not decoded yet.
        \EndIf
        \State \textbf{Stage 2:} Acknowledgment feedback
        \If{accumulative feedback is supported}
            \State Send back the index of the first information packet that is not decoded yet and its associated counter.
        \Else
            \State Send back ACKs for the received RLNC packets.
        \EndIf
        \Statex\textbf{\underline{There are changes in the network}}
        \If{link rates change}
        \ForEach{involved VN influenced by the link rate changes}
            \State Match the incoming and outgoing links of each node in the VN using Algorithm~\ref{alg:Balancing}
            \State Identify the global paths and their rates for applications/services in the FT using Algorithm \ref{alg:global_paths}.
        \EndFor        
        \ForEach{involved Source Net node}
            \State Identify the global paths and their rates for applications/services in the FT using Algorithm~\ref{alg:global_paths}.
        \EndFor 
        
        \EndIf
        \If{a node leaves or a joins the network}
            \State Call Main-Controller to update the RT.
            \State For services with updated RTs and FTs use the initialization steps given in Algorithm \ref{alg:main}. 
        \EndIf
    \EndWhile
  \end{algorithmic}
\end{algorithm}

\begin{itemize}
    \item \textbf{User:} User needs to provide information packets related to the application/service to the source nodes (Source Enc and Net nodes). If User does not have the same IP address as the source nodes, it uses traditional methods to encode information packets into IP packets and transmit them to Enc Node (lines~2-4).

    \item \textbf{Source Enc and Net nodes:} Source nodes do their roles in the AC-RLNC solution in three stages: At stage 1 (line 5), the budgeting module for the current application/service is called to distribute the global paths between new-transmissions (type~1) and re-transmissions (type~2). The budgeting module performs a bit-filling optimization, using Algorithm~\ref{alg:Budgeting}, to perform this optimization. At stage 2 (line 6), the encoding module prepares the RLNC packet (a collection of re-transmissions and new transmissions). At stage 3 (line 7), the packet allocation module embeds each RLNC packet into an IP packet and assigns it to an appropriate global path (its corresponding IP address).

    \item \textbf{VN Net node:} Each VN Net node calls the budgeting module for the current application/service to distribute the global paths in the VN between new-transmissions (type~1) and re-transmissions (type~2), line 8.

    \item \textbf{ReEnc and Relay nodes:} For each VN Relay node, the incoming RLNC packets are forwarded on appropriate global paths using the RT (lines~9-11). VN ReEnc nodes do their roles in the AC-RLNC solution in two stages (lines~12-15): At stage 1, the re-encoding module is called to prepare NEW-RLNC packets and REP-RLNC packets. More detail on re-encoding module is given in Algorithm~\ref{alg:Forward}. At second stage, the packet allocation module embeds each RLNC packet into an IP packet and assigns it to an appropriate global path.

    \item \textbf{Dec node:} Dec node performs its tasks in two stages: decoding and transmission of feedback. At stage 1, lines~16-24, the decoding module is called that checks if sufficient number of RLNC packets are received to a recover an information packet. If true, the related RLNC packets are decoded, e.g., using Gaussian elimination method. Dec node needs to provide the retrieved information packet(s) to Destination. If IP addresses of Dec node and Destination are not the same, the information packets are encoded into IP packets and transmitted via traditional methods to Destination. When the RLNC packets received so far are not sufficient to recover new information packets, the Dec at this stage just records the number of received AC-RLNC packets that can help toward decoding first information  packet that is not decoded  yet (this index is denoted by $w_{min}$, and the recorded number is denoted by Dof. At Stage 2, lines~25-30, Dec node prepares and send the feedback messages. If cumulative feedback feature is supported, $w_{min}$ and Dof are transmitted as feedback that convey the information about all packets that are decoded in-order at Destination and also the number of received RLNC packets toward decoding the first information packet that has not yet been decoded at Destination. In case the cumulative feedback feature is not supported, Dec Node sends an ACKs for all received RLNC packets.
\end{itemize}

At each time step, the SSE-SDNC checks if there are important changes in the network that affects the current application/service, e.g., changes in link rates, network nodes, etc, and performs necessary updates (lines~31-44). If link rates have notable changes, the following tasks are performed: the incoming and outgoing links of each node in the involved VNs are matched again using Algorithm~\ref{alg:Balancing}. Then, the global paths and their rates for the involved Net nodes are identified again using Algorithm~\ref{alg:global_paths}. For tracking link rate changes, a threshold function can be defined such that if the temporal link rate surpasses the threshold, the change is considered notable. For example, the threshold parameter can be set to $1.5$ times the standard deviation of the link rate, and the temporal rate can be calculated during a specific period such as $3$RTT time step. There is also a possibility that the architecture of the network changes during data transmission, i.e., an influential node joins or leaves the network. Then, the main controller needs to update the RT and perform the subsequent initialization using Algorithm~\ref{alg:main}. 

\subsection{Implementation of AC-RLNC Modules}\label{sec_modules_implementation}

\begin{figure}
    \centering
    \begin{tikzpicture}[align=center,node distance=4cm,>=stealth',bend angle=45,auto]

\tikzstyle{layers}=[rectangle,thick,draw=gray!75,fill=gray!15,minimum width=2.5cm,minimum height = 7.6cm,node distance=2.7cm]

\tikzstyle{layer}=[rectangle,thick,draw=gray!75,fill=gray!25,minimum width = 2.3cm,minimum height = 1.4cm,,text depth = 0.65 cm]

\tikzstyle{module}=[rectangle,thick,draw=black,fill=gray!40,minimum width = 1.0cm,minimum height = 0.6cm]

\begin{scope}
    
    
    \node [layers,draw,label={north:\small User/Dest/Relay}] (others) {};
    
    \node [layers,draw,label={north:\small Enc Node}] (enc) [right of=others] {};
    
    \node [layers,draw,label={north:\small Source Net Node}] (srcNet) [right of=enc] {};
    
    \node [layers,draw,label={north:\small VN Net Node}] (vnNet) [right of=srcNet] {};

    \node [layers,draw,label={north:\small ReEnc Node}] (ReEnc) [right of=vnNet] {};
    
    \node [layers,draw,label={north:\small Dec Node}] (dec) [right of=ReEnc] {};
    
    \foreach \x in {0,2.7,8.1,10.8} {
    
        \node [layer,draw] (Application) at(\x,3.0) {\small Application};
        
        \node [layer,draw] (TCP) at(\x,1.5) {\small TCP};
    
        \node [layer,draw] (IP) at(\x,0) {\small IP};
    
        \node [layer,draw] (Network) at(\x,-1.5) {\small Network};
    
        \node [layer,draw] (Physical) at(\x,-3.0) {\small Physical};
    }
    
    \foreach \x in {5.4,13.5} {
    
        \node [layer,draw] (Application) at(\x,3.0) {\small Application};
        
        \node [layer,draw] (TCP) at(\x,1.5) {\small MPTCP};
    
        \node [layer,draw] (IP) at(\x,0) {\small IP};
    
        \node [layer,draw] (Network) at(\x,-1.5) {\small Network};
    
        \node [layer,draw] (Physical) at(\x,-3.0) {\small Physical};
    }
    
    
    \node [module,draw] (encoding) at(2.7,2.7) {\footnotesize ENC};
    
    \node [module,draw] (gpi) at(4.85,2.7) {\footnotesize GP};
    \node [module,draw] (gpi) at(7.55,2.7) {\footnotesize GP};
    
    \node [module,draw] (gpi1) at(5.95,2.7) {\footnotesize BUG};
    \node [module,draw] (gpi2) at(8.65,2.7) {\footnotesize BUG};
    
    \node [module,draw] (allocation) at(5.4,1.2) {\footnotesize PA};
    \node [module,draw] (allocation) at(10.8,-0.3) {\footnotesize PA};
    
    \node [module,draw] (balancing) at(10.25,2.7) {\footnotesize BAL};
    
    \node [module,draw] (reencoding) at(11.35,2.7) {\footnotesize REC};
    
    \node [module,draw] (decoding) at(13.5,2.7) {\footnotesize DEC};

\end{scope}

\end{tikzpicture}
    \caption{Implementation of AC-RLNC modules.}
    \label{fig:networklayers}
\end{figure}
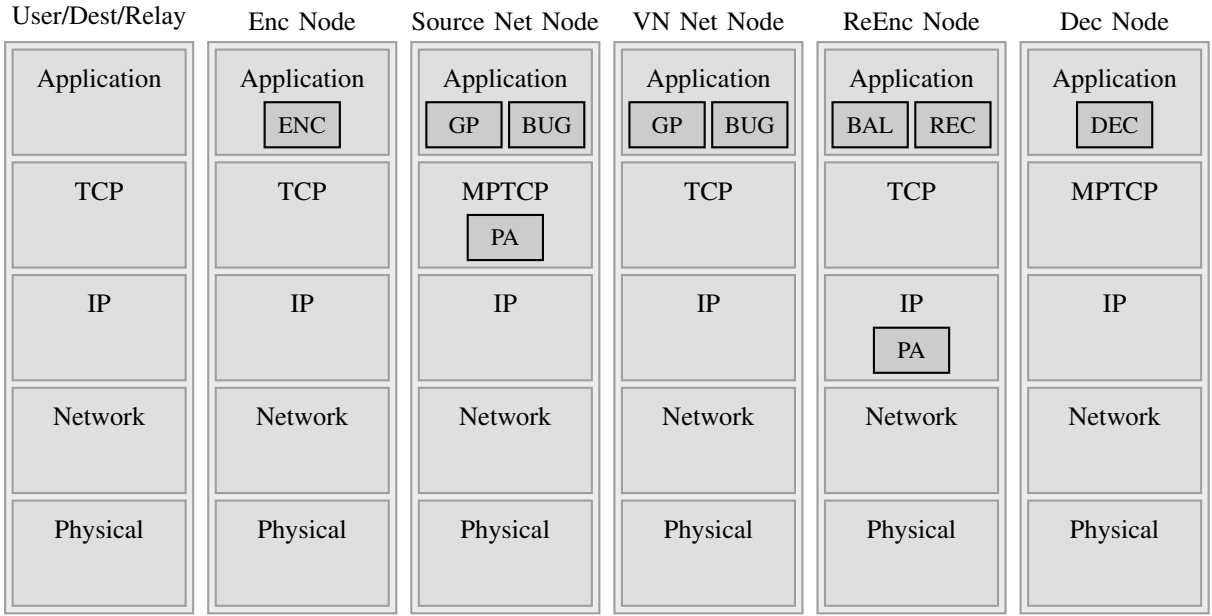

In this subsection, we provide algorithm for each AC-RLNC module. Then, we present in which network layer of the Open Systems Interconnection model (OSI model) each module needs to be implemented based on the information the module needs and the nature of its algorithm. Fig.~\ref{fig:networklayers} shows the proposed network OSI stack layer implementation for the AC-RLNC solution.

\subsubsection{Balancing Module Implementation}

We remind that the role of balancing module is matching the incoming links and outgoing links of ReEnc nodes in a VN to improve the transmission throughput. The balancing module does this task by matching the incoming and outgoing links such that the link rate variations and correspondingly the bottleneck effect is minimized over VN global paths. For this purpose, a so called \textit{natural matching} is defined in \cite[Theoem 3]{cohen2020adaptiveMPMH} and briefly reviewed here: Consider the links between any two nodes in a VN are sorted and indexed in rate-decreasing order. Natural matching is matching the $p$'th incoming link with $p$'th outgoing link ($p\in\{1,\dots,P\}$ and $P$ is the number of global-paths). It is shown in \cite{cohen2020adaptiveMPMH} that the natural matching is the optimal matching in terms of the resultant VN transmission throughput.

In the balancing procedure introduced in \cite{cohen2020adaptiveMPMH}, it was assumed all hops (intermediate nodes) are capable of performing the balancing. However, in this paper, we consider a more general case where there can exist one or more Relay nodes, not capable of running the balancing module, between two consecutive ReEnc nodes in a VN. This extension which provides the heterogeneity property in the presented architecture necessities new considerations in the balancing module. We also assume the first node and the last node in each VN are ReEnc nodes. In matching the incoming links and outgoing links at a ReEnc node, the rate of each incoming and outgoing links are replaced with so called \textit{associated rate}. The associated rate of an incoming link (resp., outgoing link) is summation of link rates on the global path that the link belongs to, starting from the previous ReEnc node (resp., ending to the next ReEnc node). In this way, the balancing module takes into account that the Relay nodes are not capable of performing link matching and makes more informed decisions.

\begin{figure}
    \centering
    \includegraphics[trim=0cm 0.0cm 0cm 0cm,width = 1 \columnwidth]{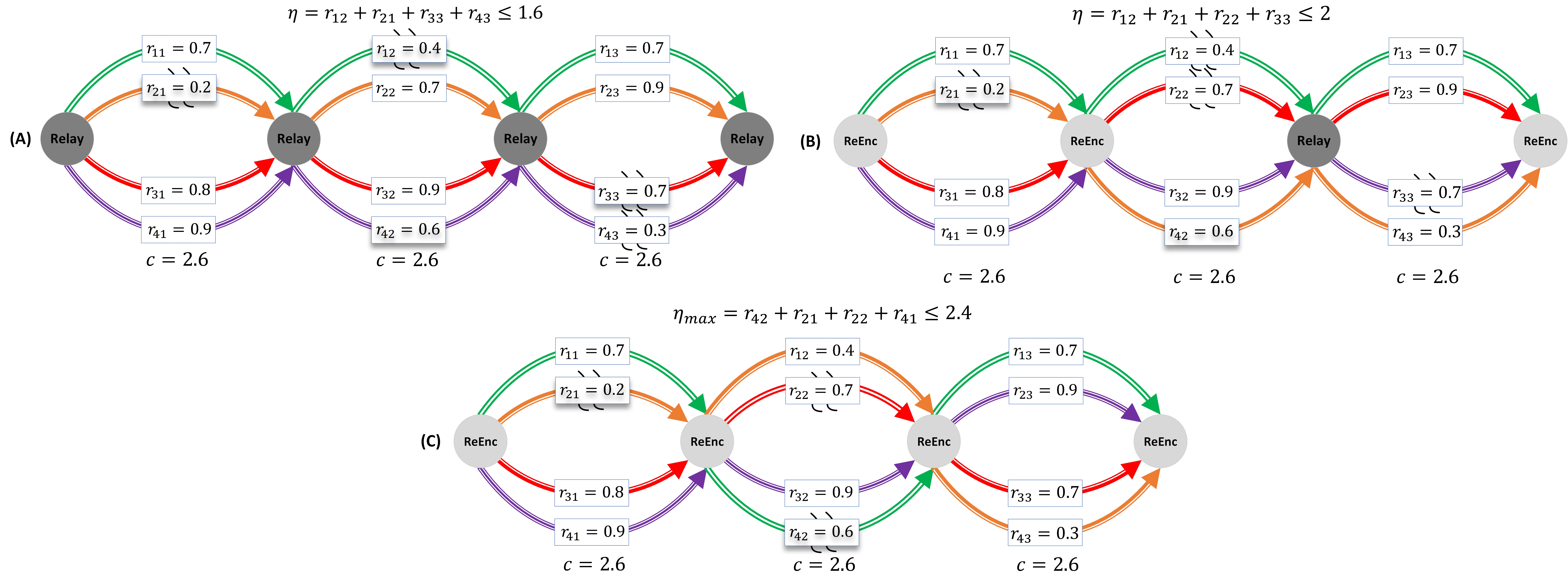}
    \caption{Matching between input and output links in a VN. (A) shows the matching between input and output links for a naive choice of global paths. (B) shows the matching between input and output links when there is one intermediate node that is just Relay. (C) shows the matching between input and output links when all nodes are ReEnc nodes.}
    \label{fig_matching_opt}
\end{figure}

Fig.~\ref{fig_matching_opt} show an example of matching between input and output links in three different scenarios. The capacity between any two consecutive nodes is $2.6$ in this example (summation of link rates between the two nodes). Each color represents one global path, and the rate of each global path is determined by the rate of the constituent link with lowest rate (marked in case the constituent links do not have a fixed rate and there is a bottleneck). The throughput then is summation of the rate of global paths. In scenario (A), a naive choice of global paths is considered because all intermediate nodes are Relay nodes and are not capable of optimal matching. The throughput in this scenario is $1.6$. In scenario (B), a semi-optimal choice of global paths is considered. Here, one intermediate node is ReEnc node and capable of natural ordering of its incoming links. However, the second intermediate node is Relay node and is not capable of natural ordering. Thus, the last ReEnc node performs the natural ordering of its incoming links using their associated rates to take into account that the previous Relay node only does a native matching. The throughput in this scenario is $2.0$. In scenario (C), an optimal choice of global paths is considered. Here, all nodes are ReEnc nodes and capable of natural ordering of their incoming links. The throughput in this scenario is $2.4$.

The balancing procedure can be implemented in a distributed fashion such that each ReEnc node in a VN performs the natural ordering of its incoming links based on their associated rates and sends the results to the next ReEnc node. To attain the associated rates for the incoming links, each ReEnc node requires the routing information and incoming/outgoing link rates of the relay nodes that appear between the previous ReEnc node and itself. When the natural ordering of incoming links are performed consecutively at each ReEnc node, in order they appear in the VN, and the results are sent to the next ReEnc node, each ReEnc is able to update the LPRT for the previous ReEnc node. If the Relay nodes do not supply feedback information to the next ReEnc node to estimate the associated link rates, the main controller needs to provide this information to the ReEnc nodes. In the initialization process in the Main controller, the balancing module at each ReEnc node in a VN requests the RT of previous Relay nodes (up to previous ReEnc node). In case of changes in the RT of the Relay node, the main controller inform the first ReEnc node after the Relay node about the change.

As algorithm~\ref{alg:Balancing} presents, the balancing module for a VN receives from the controller the incoming and outgoing associated rates for the current application/service. Then, for each ReEnc node, in order they appear in the VN (see Fig.~\ref{fig_MP_MH_in_Bigger_Net}), it finds the optimal matching and updates its LPRT accordingly. Because of the algorithmic nature of the balancing module, it is better to be implemented in application layer of each VN Re-Enc node.

\begin{algorithm}\small 
  \caption{Balancing module \cite{cohen2020adaptiveMPMH}}
  \label{alg:Balancing}
  \begin{algorithmic}[1] 
    \Statex\textbf{Input:} Input and output links' associated rates for each ReEnc node in a VN.
    \ForEach{ReEnc node in order they appear in the VN}
        \State Match the incoming and outgoing links to the node using the natural matching.
        \State Send the matched order of the output links to the next ReEnc node in the VN.
    \EndFor
    \Statex\textbf{Output:} Update LPRT      
  \end{algorithmic}
\end{algorithm}

\subsubsection{Global-Path Identification Module Implementation}

We remind  that  the role of global path identification  module is to find all global paths along with their rates between two parts of the network architecture for an application/service. If the global path identification module is called by a Source Net node, the two parts are Source Enc and Dest Dec nodes. If the global path identification module is called by a VN Net node, the two parts are the first and last nodes in the VN.

Algorithm~\ref{alg:global_paths} is given for the global path identification module. The module called by a Source Net node uses the GPRTs associated with the involved VN Net nodes for a faster global path identification from Source to Destination. For a VN, a naive choice of global paths is attained by arbitrarily matching the input and output links of the constituent nodes. A naive choice of global-paths is used when a new application/service uses the VN. An optimal choice of global-paths is used for an existing application/service as LPRTs already exist for the ReEnc nodes. Then, the rate for each global path ( $r_{G_p}$ for $p$-'th global path, $p\in\{1,\dots,P\}$ and $P$ is the number of global paths) is computed and recorded in the GPRT. Global-path identification module is implemented in application layer of each Net node.

\begin{algorithm}\small 
    \caption{Global path identification module}
    \label{alg:global_paths}
    \begin{algorithmic}[1] 
    \Statex\textbf{Input:} RTs, FTs, GPRTs, LPRTs.
    \If{Called by a Source Net Node}
        \State Create GPRT using GPRTs of associated VN Net nodes for the current application/service.
    \Else
        \If{New service/application}
        \State Create GPRT with naive choice of global paths.
    \Else
        \State Update GPRT with the updates LPRTs.
    \EndIf
    \EndIf
    \State Calculate the rate for each global path in GPRT.
    
    \Statex\textbf{Output:} Send rate of all global paths to the budgeting module as input.
    
  \end{algorithmic}
\end{algorithm}

\subsubsection{Budgeting Module Implementation}

This module is used in two types of nodes in the solution proposed, Source Net nodes and VN Net nodes. First, we will show the role of budgeting  module at the Source Net nodes, and then we will explain the differences in the implementation of this module at the VN Net nodes. The role of budgeting module is to determine, at each time slot $t$, which global paths are of type~1, i.e., their number denoted by $N_\text{new}$, and which are of type~2, their number denoted by $N_\text{ret}$. At the Source Net nodes, it also keeps track of the effective window limits at each time step, i.e., $w_{min}$ and $w$, requests the encoding module to prepare $N_\text{new}$ NEW-RLNC packets and $N_\text{ret}$ REP-RLNC packets at each time steps, and updates window limits accordingly. We note that the feedback is received from Dec node (resp., last ReEnc node) for the current application/service in the network (resp., the VN), and thus the budgeting module is aware of the received RLNC packets at the designated node after some delay.

We now review some necessary notations. The RLNC packet that is sent at time slot $t$ over the $p$'th global path is denoted by $c(t,p)$, which is either a re-transmission or a new-transmission. The parameter $w_{min}$ and $w$ are the beginning and the length (in terms of the number of packets) of the effective window of the information packets. The parameter $r_{G_p}$ denotes the rate of $p$'th global path and is updated during data transmission. The parameters $m_{d_g}$ and $a_{d_g}$ are the number of missing degrees of freedom and added degrees of freedom for decoding the information packets in the current effective window. The re-transmission criterion is defined as $\Delta>0$, which will be described more in sequel. The parameter $\overline{o}$ is referred to maximum effective window size, i.e., $w\leq \overline{o}$. The parameter $m_{G_p}$ denotes the number of FECs that are transmitted over $p$'th global path, per generation of $k$ raw information packets (A reasonable choice is $k=RTT-1$). The FEC packets are in fact a subset of re-transmissions that compensate in advance for the erasures that occur over a lossy channel regardless of the feedback\footnote{For example, when $p$'th global path can be modeled as a BEC with parameter $\epsilon_{p}$, $m_{G_p}$ can be set to $\ceil{ \epsilon_{p}k}$.}. For $p$'th global path and after transmitting first $k$ new-transmissions, re-transmissions are sent in next $m_{G_p}$ time slots, as FECs. The remaining re-transmission packets that are not FEC and are sent in response to the received feedbacks are called FB-FECs.

At source Net nodes and at each time slot $t$, the budgeting module is run. First, it checks if feedback is available from Dec node. If no feedback is available yet (occurs in first $k$ time steps or in case of feedback reception failure), all global paths are considered for new-transmissions (have type~1) as long as the end of generation of new packets (EW event) is not reached. If EW occurs and no feedback is received, all global paths are considered for re-transmissions (have type~2), lines~1-7.
If feedback is available, the module makes necessary updates, lines~6-12. Based on the erasure patterns, the rate of each global path, i.e., $r_{G_p}$, is updated. Next, $w_{min}$ is incremented by the number of new information packets that the decoding module has managed to decode in order at time slot $t-RTT$, and $w$ using the updated $w_{min}$ is updates as $w=w_{max}-w_{min}$. Then, the number of missing degrees of freedom $m_{d_g}$ and the number of added degrees of freedom $a_{d_g}$ are updated\footnote{For an example of how to update the parameters $m_{d_g}$ and $a_{d_g}$, refer to \cite{cohen2020adaptiveMPMH}}. We remind that the parameters $m_{d_g}$ is the number of new-transmissions that are erased, and $a_{d_g}$ is the number of re-transmissions that are received at Dec node, for decoding the information packets in the current effective window. The budgeting module balances the new-transmissions and re-transmissions to hold $m_{d_g}<a_{d_g}$. The parameter $\Delta$ is used to provide a desired trade-off between the throughput and delay. For a desired trade-off between throughput and in-order delivery delay, a tune-able margin $th$ is considered, and the re-transmission criterion is defined as $\Delta=P.(m_{d_g}/a_{d_g}-1-th)>0$.

After necessary updates based on the received feedback, the budgeting module checks whether the maximum effective window size is reached. If true, all global paths are used for re-transmissions, lines~13-15. Otherwise, the budgeting modules effectively identifies the global paths that will carry new-transmissions and global paths that carry re-transmission. The parameter $m_{G_p}$ shows the number of FECs that is due for the $p$'th global path as mentioned before. For FEC re-transmissions (lines~17-21), each global path $p$ with $m_{G_p}>0$ is selected for re-transmission, and $N_\text{ret}$ and $m_{G_p}$ are adjusted accordingly. For remaining global paths, the module identifies the global paths for FB-FEC re-transmissions, and selects the rest for new-transmissions. How the budgeting module does this split is based on a bit-filling optimization that maximizes the throughput of new information packets while minimizing the in-order delivery delay by providing sufficient re-transmissions. This optimization problem was introduced in \cite{cohen2020adaptiveMPMH}, and we revisit it here in Proposition~\ref{PDWF1}. In fact, if $\Delta>0$, FB-FEC re-transmissions are needed and the appropriate global paths (of type~2) are selected according to the Proposition~\ref{PDWF1}, and $N_\text{ret}$ is adjusted accordingly, lines~13-27. Then, the remaining paths (if any) are considered for new-transmissions in case there are still new information packets to be added to the effective window, and $N_\text{new}$ is adjusted, lines~28-33.
When the end of generation of new packets is reached ($EW$ event), the initialization for FEC re-transmissions are performed for all global paths, and also the unassigned global paths at the current time step are selected for FEC re-transmissions and $N_\text{new}$ and their $m_{G_p}$s are adjusted accordingly, lines~34-41. Finally, the budgeting module requests $N_\text{new}$ NEW-RLNC packets and $N_\text{ret}$ REP-RLNC packets from the encoding module, updates $w$ based on the new-transmissions occurred in the current time step, and sends the RLNC packets along with the global paths' types to the packet allocation module, lines~45-46. Algorithm~\ref{alg:Budgeting} describes all the aforementioned steps.

For the bit-filling optimization solution as given in \cite{cohen2020adaptiveMPMH}, we define the set of available global paths between the source to the destination as $\mathcal{P}$, and the index of a possible sub-set as $\xi\in\{1,\ldots,2^{P}\}$. We denote the global paths with indices in $\xi$ as $\mathcal{P}_\xi$, and   $\mathcal{P}_{\xi}^c=P\setminus \mathcal{P}_{\xi}$.
\begin{prop}[Bit-Filling \cite{cohen2020adaptiveMPMH}]\label{PDWF1}
Given the estimated rates of available global paths between the source and the destination, $r_{G_p}$ and $p\in\{1,\ldots,P\}$, the source intends to maximize the throughput of the new packets of information. The set of paths $\mathcal{P}_{\hat\xi}$ on which new packets are sent, i.e., have type~1, is obtained as
\begin{equation}
\label{DWF}
\hat\xi = \underset{\xi}{\arg\max} \sum_{i\in \mathcal{P}_\xi} r_{G_i}, \quad \text{s.t.} \quad \sum_{j\in \mathcal{P}_{\xi}^c} r_{G_j} \geq \Delta \quad 
\end{equation}
where the optimization problem minimizes the in-order delivery delay, by providing over the type~2 global paths a sufficient number of DoF's for decoding.
\end{prop}

Now we elaborate on the role of the budgeting module at the VN Net nodes. The topology in each VN can be different from the general topology of the network. Namely, the number of global paths at the VNs can differ from the available global paths from source to destination. To increase the efficiency in the VNs, in terms of throughput-delay tread-off, this module determines at each time step $t$, out of the $P$ global paths at the specific VN, which are of type~1 and which are of type~2. However, unlike the implementation of this module at the Source node, it is not required to manage the window size, as new raw information packets are not available at VN Net nodes. A new bit-filling optimization is given in Proposition \ref{PDWF} to determinate types of the global paths at VNs, and then this information is provided to the ReEnc node (see Section \ref{subsec:Re-Encoding}) to utilize the different types of packets at the incoming links to re-encode and allocate them according to the determination of the budgeting module. In case there is no incoming NEW-RLNC packets but the bit-filling optimization results in $N_\text{new}>0$, the module sets $N_\text{ret}=P$. Note that the budgeting module is used only when the selective mixing is supported in the VN.

We denote the set of available incoming paths to and constituent paths of the VN Net as $\mathcal{P}(s)$, where $s$ distinguishes between incoming paths and outgoing paths respectively, i.e., $s\in\{In,Out\}$.
We denote the index of a possible sub-set as $\xi(s)\in\{1,\ldots,2^{P(s)}\}$, and the paths with indices in $\xi(s)$ as $\mathcal{P}_\xi(s)$. Again, $\mathcal{P}_\xi^c(s)=P(s)\setminus \mathcal{P}_\xi(s)$. Here, $\{p|p\in\mathcal{P}_{\hat\xi}^c(In)\}$ represents the incoming global paths to the VN that have type~2 (carry re-transmissions).

\begin{prop}[VN Net Bit-Filling]\label{PDWF}
Given the estimated rates of the incoming global paths to the VN, i.e., $\{r_{G_p}(In)|p\in\{\text{incoming global path}\}\}$, and the estimated rates of the VN global paths, i.e., $\{r_{G_p}(Out)|p\in\{\text{VN global path}\}\}$, the budgeting module intends to maximize the throughput of the NEW-RLNC packets in the VN. The set of paths $\mathcal{P}_{\hat{\xi}}(Out)$ on which NEW-RLNC packets are sent, i.e., have type~1, is obtained as
\begin{equation}
\begin{aligned}
\hat{\xi}(out) = \underset{\xi(Out)}{\arg\max}
\sum_{i\in \mathcal{P}_\xi(Out)} r_{G_i}(Out), \quad \text{s.t.} \quad \sum_{j\in \mathcal{P}_{{\xi}}^c(Out)} r_{G_j}(Out) \geq \sum_{l\in \mathcal{P}_{\hat\xi}^c(In)} r_{G_l}(In) \quad  
\end{aligned}
\end{equation}
where the optimization problem minimizes the in-order delivery delay, by providing over the type 2 global paths a sufficient number of DoF’s at the VN.
\end{prop}

The new bit-filling optimization in Proposition~\ref{PDWF} provides the opportunity to handle the heterogeneity inside a VN network, e.g., having a variant number of paths between two consecutive intermediate nodes. In this case, a Net node needs to be associated with each ReEnc node (not just the first one as depicted in Fig~\ref{fig_MP_MH}) to perform the budgeting module (distribution of new-transmissions and re-transmissions along the outgoing links using the optimization described in Proposition~\ref{PDWF}) such that both throughput maximization and  delivery delay minimization targets are satisfied.

\begin{algorithm}\small
\caption{Budgeting module at Source Net nodes \cite{cohen2020adaptiveMPMH}}
\label{alg:Budgeting}
\begin{algorithmic}[1]
\Statex\textbf{Input:} Feedback acknowledgment, $r_{G_p}$, rates and unencoded RLNC data.  
\Statex\textbf{Init:} $N_\text{new}=0$ and $N_\text{ret}=0$

        \State 
        \If{no feedback available}
            \If{EW}
                \State $N_\text{ret}=P$ and $N_\text{new}=0$
            \Else
                \State $N_\text{ret}=0$ and $N_\text{new}=P$
            \EndIf
        \Else
            \State Update $r_{G_p}$ for each path
            \State Update $w_{min}$ and $w$
            \State Update $md_g$ and $ad_g$
            \State Update $\Delta$
            \State \textbf{Size limit re-transmissions:}
            \If{$w >\bar{o}$}
                \State  $N_\text{ret}=P$, $N_\text{new}=0$
            \Else
                \State \textbf{FEC re-transmissions:}
                \ForEach{global-path $p$ with $m_{G_p}>0$}
                    \State $N_\text{ret}= N_\text{ret}+1 $
                    \State $m_{G_p} = m_{G_p}-1$
                \EndFor
            \If{$P-N_\text{ret}>0$}
                \State \textbf{FB-FEC re-transmissions:}
                \If{$\Delta>0$}
                    \State Determine FB-FEC paths (Proposition \ref{PDWF1})
                    \State $N_\text{ret}=N_\text{ret}+\sum \text{FB-FEC paths}$
                \EndIf
                \State \textbf{new-transmissions:}
                \ForAll{remaining $P-N_\text{ret}-N_\text{new}$ paths}
                    \If{not EW}
                        \State $N_\text{new}=N_\text{new}+1$
                    \EndIf
                \EndFor
                \State \textbf{FEC re-transmissions (initialization):}
                \If{EW}
                    \State Set $m_{G_p} := \lceil \epsilon_{G_p} (RTT-1)\rfloor$
                    \ForAll{remaining $P-N_\text{ret}-N_\text{new}$ paths}
                        \State $N_\text{ret} = N_\text{ret} + 1$
                        \State $m_{G_p} = m_{G_p}-1$
                    \EndFor
                \EndIf
            \EndIf
        \EndIf
    \EndIf
\State Request from Enc module $N_\text{new}$ NEW-RLNC and $N_\text{ret}$ REP-RLNC packets with $w_{min}$ and $w$
\State $w = w + N_\text{new}$
\Statex\textbf{Output:} Send the RLNC packets to the Packet Allocation module
\end{algorithmic}
\end{algorithm}

\subsubsection{Encoding Module Implementation}

\begin{algorithm}\small 
  \caption{Encoding module \cite{cohen2020adaptiveMPMH}}
  \label{alg:Encoding}
  \begin{algorithmic}[1] 
  \Statex\textbf{Input:} $N_\text{new}$ and $N_\text{ret}$, $w_{min}(t)$ and $w(t)$
  
    \If{User and Source nodes (Source Enc and Net nodes) do not have same IP address}
        \State Decode the IP packets received from the User.
    \EndIf
    
    \ForRange{$N_\text{ret}$}
        \State Draw $w(t)$ random coefficients for encoding process, $\{\mu_{1}, \ldots, \mu_{w(t)}\}$. \cite{pedersen2011kodo}.
        \State Pick information packets $\{p_i| i\in \{w_{min}(t),\dots, w_{min}(t)+w(t)\}\}$. 
        \State Prepare re-transmissions: $c(t,k)=\sum_{i=w_\text{min}(t)}^{w_\text{min}(t)+w(t)}\mu_ip_i$ 
    \EndFor
    \ForRange{$N_\text{new}$} 
        \State $w_k = w(t) + k$
        \State Draw $w_k$ random coefficients for encoding, $\{\mu_{1}, \ldots, \mu_{w_k}\}$.
        \State Pick information packets $\{p_i| i\in \{w_{min}(t),\dots, w_{min}(t)+w_k\}\}$.  
        \State Prepare new-transmissions: $c(t,k+N_{ret})=\sum_{i=w_\text{min}(t)}^{w_\text{min}(t)+ w_k}\mu_ip_i$
    \EndFor
    \Statex\textbf{Output:} $N_\text{new}+N_\text{ret}$ RLNC packets.
  \end{algorithmic}
\end{algorithm}

The  role of  encoding  module  is preparing the RLNC packets, Algorithm~\ref{alg:Encoding}. The module first retrieves the application/service information packets from the received IP packets in case User and Source nodes do not have the same IP addresses (lines 1-3). An RLNC packet at time step $t$ is formed by random linear combination of the application/service information packets in a properly defined sliding window. If for an RLNC packet $w_{max}(t)=w_{max}(t-1)$, it is a re-transmission and denoted by REP-RLNC. Otherwise ($w_{max}(t)>w_{max}(t-1)$), it is called a new-transmission and is denoted by NEW-RLNC. At time step $t$, the encoding module prepares $N_\text{ret}$ Re-RLNC packets (lines~4-8) and  $N_\text{new}$ New-RLNC packets (lines~9-14) and send all packets to the packet allocation module as input. We note that $N_\text{new}+N_\text{ret}=P$ is number of global paths where $N_\text{new}$ is the number of global paths of type~1 and $N_\text{ret}$ is the number of global paths of type~2). Encoding module is implemented in application layer of each Source Enc node.

\subsubsection{Re-Encoding Module Implementation}\label{subsec:Re-Encoding}

The role of re-encoding module is preparing new set of RLNC packets at ReEnc nodes, and is described in Algorithm~\ref{alg:Forward}. If selective mixing is supported, the incoming NEW-RLNC packets received at the current time step are mixed together (via re-encoding using a new set of random coefficients), line~3, and the incoming REP-RLNC packets received at current time step and previous time steps (though buffers) are mixed together, line~4. If traditional mixing is supported, the incoming RLNC packets received at current time step and previous time steps (though buffers) are all mixed together, line~6. If no mixing is supported, received RLNC packet from each incoming link is mixed with RLNC packets received from the same link at previous time steps, line~8. Re-encoding module is implemented in application layer of each VN Re-Enc node. If the NEW-RLNC packets (resp., REP-RLNC packets) packets received at a ReEnc node are fewer than the number of global paths of type~1 (resp., global paths of type~2), by re-encoding over the received NEW-RLNC packets (resp., REP-RLNC packets) the missing packets are compensated.

\begin{algorithm}\small 
    \caption{Re-encoding module \cite{cohen2020adaptiveMPMH}}
    \label{alg:Forward}
    \begin{algorithmic}[1] 
    \Statex\textbf{Input:} Received RLNC packets (NEW-RLNC packets and REP-RLNC packets).
    \State Draw random coefficients for re-encoding process.
    \If{Selective-mixing is supported}
        \State Re-encode received packets at time slot $t$ from global paths of type 1  (with new random coefficients).
        \State Re-encode received packets at time slots $\leq t$ from global paths of type~2.
    \ElsIf{Traditional mixing is supported}
        \State Re-encode received packets at time slots $\leq t$ from all global paths. 
    \Else
        \State Re-encode independently received packets at time slots $\leq t$ from each global paths.
    \EndIf
    \Statex\textbf{Output:} Re-encoded RLNC packets
  \end{algorithmic}
\end{algorithm}

\subsubsection{Packet Allocation Module Implementation} The role of this module is embedding the given RLNC packets into IP packets. If an IP packet includes a NEW-RLNC packet, the module sends it through a global path of type 1 (lines~2-5); otherwise, if it includes a REP-RLNC packet, the module sends it through a global path of type 2 (lines~6-9). Packet Allocation module is implemented in network layer of Source Net nodes and VN Re-Enc nodes.

\begin{algorithm}\small 
  \caption{Packet Allocation module}
  \label{alg:PacketAllocation}
  \begin{algorithmic}[1]
    \Statex\textbf{Input:} $P$ RLNC packets, GPRT, $w_{min}(t)$, $w_(t)$ and $\vec{\mu}$.
    \ForEach{RLNC packets}
        \If{NEW-RLNC packet}
            \State REP = 0.
            \State Encode the RLNC packet in IP packet using the network information showed in Fig.~\ref{fig:packet}.
            \State Allocate the IP packet to global path of type 1
        \Else
            \State REP = 1.
            \State Encode the RLNC packet in IP packet using the network information showed in Fig.~\ref{fig:packet}.
            \State Allocate the IP packet to global path of type 2
            \EndIf
    \EndFor 
    \Statex\textbf{Output:} Transmit $P$ IP packets through appropriate global paths
  \end{algorithmic}
\end{algorithm}

\begin{figure}
    \centering
    \begin{tabular}{c}
    \begin{tikzpicture}[align=center,node distance=4cm,>=stealth',bend angle=45,auto]

\begin{scope}
    \node [rectangle,draw,thick,draw=gray!75,fill=gray!20,minimum width=6cm,minimum height = 1.5cm] (data) {Data};
    \node [rectangle,draw,thick,draw=gray!75,fill=gray!20,minimum width = 4cm,minimum height = 1.5cm,node distance=5.0cm] (header) [left of=data] {TCP header};
    
\end{scope}

\end{tikzpicture}\\
    (a)\\
    \begin{tikzpicture}[align=center,>=stealth',bend angle=45,auto]

\begin{scope}
    \node [rectangle,draw,thick,draw=gray!75,fill=gray!20,minimum width=3.5cm,minimum height = 1.5cm,align=center] (RLNC) {RLNC packet};
    \node [rectangle,draw,thick,draw=gray!75,fill=gray!20,minimum width = 1cm,minimum height = 1.5cm,node distance=2.25cm] (coeff) [left of=RLNC] {$\bf{\vv{\mu}}$};
    \node [rectangle,draw,thick,draw=gray!75,fill=gray!20,minimum width = 1cm,minimum height = 1.5cm,node distance=1cm] (w) [left of=coeff] {$\bf{w}$};
    \node [rectangle,draw,thick,draw=gray!75,fill=gray!20,minimum width = 1.2cm,minimum height = 1.5cm,node distance=1cm] (wmin) [left of=w] {$\bf{w_\text{min}}$};
    \node [rectangle,draw,thick,draw=gray!75,fill=gray!20,minimum width = 1.2cm,minimum height = 1.5cm,node distance=1.2cm] (rep) [left of=wmin] {REP};
    \node [rectangle,draw,thick,draw=gray!75,fill=gray!20,minimum width = 2cm,minimum height = 1.5cm,node distance=1.7cm,text width=2cm] (destTCPport) [left of=rep] {Source TCP port};
    \node [rectangle,draw,thick,draw=gray!75,fill=gray!20,minimum width = 2cm,minimum height = 1.5cm,node distance=2.15cm,text width=2cm] (srcTCPport) [left of=destTCPport] { Destination TCP port};
    \node [rectangle,draw,thick,draw=gray!75,fill=gray!20,minimum width = 1.5cm,minimum height = 1.5cm,node distance=4.1cm,text width=1.4cm] (srcIP) [left of=destTCPport] {Source IP};
    \node [rectangle,draw,thick,draw=gray!75,fill=gray!20,minimum width = 1.5cm,minimum height = 1.5cm,node distance=1.9cm,text width=2.0cm] (destIP) [left of=srcIP] {Destination IP};
\end{scope}

\end{tikzpicture}\\
    (b)
    \end{tabular}
    \caption{Proposed (a) TCP packet structure, (b) IP packet structure.}
    \label{fig:packet}
\end{figure}
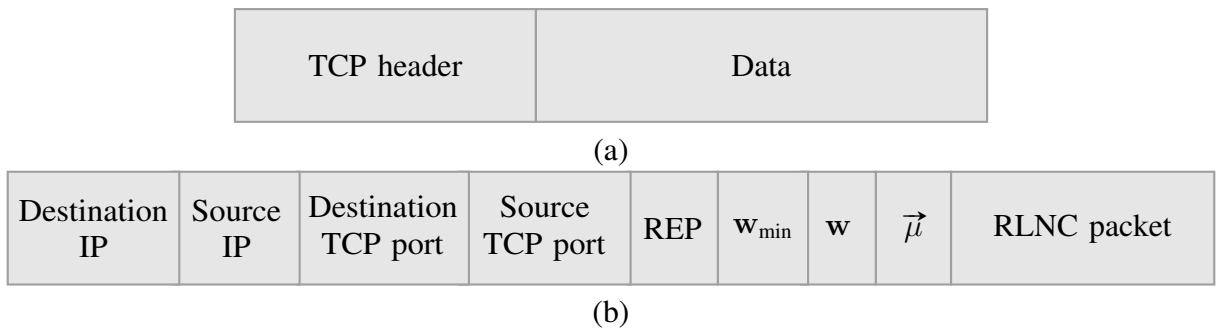

We propose standard formats for the TCP packets and the IP packets that support the AC-RLNC solution, see Fig.~\ref{fig:packet}. The information packets for the AC-RLNC solution are in fact the TCP packets\footnote{When the TCP header and the TCP paylod are not separated and are considered one information packet, the setup will be more robust against network imperfections.} \cite{sundararajan2011network,kim2012network,cloud2013multi}. Since the AC-RLNC decoding module is performed in Application layer, the source and destination TCP ports along with their IP addresses are included in the IP packet. There is also a flag in the IP packet (REP flag) that shows if embeded RLNC packet is re-transmission (REP=1) or new-transmission (REP=0). The index ($w_{min}$) of first information packet and the number information packets ($w$) that is used for preparing the RLNC packet are included in the IP packet for re-encoding and decoding purposes. Lastly, the vector of $w$ random coefficients is also embedded. We note that embedding the random coefficients in the IP packet is not necessary, and alternatively, one can embed the random seed that is used for generating the random coefficients \cite{PRNG,PRNG_Generator}.

It is possible to have more than one RLNC packets in the payload of an IP packet as suggested in \cite{sundararajan2011network,quic_fec_sof}. The benefit of incorporating more than one coded packets in an IP packet is two-fold: First, RLNC coded packets with variable lengths can be suppurated, as they will be concatenated in one IP packet with a standard length. Second, having a short RLNC code length does not impede the network performance as multiple short RLNC coded packets form an IP packet with a standard length. Having multiple RLNC coded packet in one IP packet might need extra consideration in AC-RLNC modules.

\section{Visions and enabling technologies}\label{sec:visions}

Today's technological world demands effective distributed computation and storage to handle large volume of data with growing computational needs, security requirements, and delay sensitivities. The proposed scheme in this paper that brings network coding into SDN and provides an ultra reliable communication solution for heterogeneous networks, also opens a door for invaluable technologies and features that are beyond just reliable communications. Examples are but not limited to: security feature in controller and in data transmission, mmWave use case, and distributed computation and storage. The envisioning features and technologies are described briefly in this section.

\subsection{Securing SDN controllers }
Current SDN security challenges \cite{sallam2019security,lee2020comprehensive} include proper authentication, access control, and data privacy and integrity among the orchestration components of SDN. In addition, (Distributed) Denial of Service attacks can be performed if centralized approaches for deploying SDN controllers are followed, for example due to limitations on management of flow tables that can be filled up by malicious/erroneous applications. Protection for other attacks like Man in the Middle (MITM) or replay attacks require enhancements in protocols like OpenFlow (to avoid changes in the fields’ of control messages). 

The SSE-SDNC herein proposed, tackles these security vulnerabilities by employing replication of controllers to avoid single point of failure (SPOF) and implements authentication, encryption and access control mechanisms to establish trust between the diverse components (e.g. SDN management applications, SDN controllers and network devices).
For that,  SSE-SDNC will perform replication of controllers with dynamic load distribution among controllers according to traffic conditions and load in the network, therefore addressing the SPOF issue. Authentication, access control and data privacy and integrity is achieved by integration with the Software-Defined Perimeter (SDP) security approach \cite{sallam2019security}, that employs TLS mechanisms to enable mutual authentication and encryption of communications between clients (e.g. applications, network devices) and servers (e.g. controllers). In addition, SDP features an  SDP control that is responsible by determining  which devices/applications can connect to a given component (e.g. SDN controller), thus managing access control between clients and servers. The SSE-SDNC can be instrumented to interoperate with a SDP controller to only communicate with authenticated and trustworthy devices/applications.

So far in this subsection, the security aspects for the proposed controller are elaborated. It is essential to note that the utilization of network coding with post-quantum cryptosystems, as given in \cite{cohen2020PostQuant} for data transmission, results in a very efficient and promising security safeguards for the heterogeneous communications suggested herein. Combining AC-RLNC codes with a hybrid network coding cryptosystem presented in \cite{cohen2020PostQuant} enables a strong post-quantum security level with high communication rates. Surprising, this security level can be guarantee across all the links in the MS-MD network by encrypting a single link with computational cryptography.

\subsection{mmWave mesh paths in AC-RLNC and SDN}
The MP-MH AC-RLNC approach requires the knowledge of the throughput and delay of the several links and global paths, so that global paths of type 1 and type 2 can be determined. In a radio-based communication, the throughput and delay may depend on several different parameters beyond the link capacity and the data rate, since wireless links are very prone to interference and distance. The uncertainty on these global paths’ parameters is increased if we consider the characteristics of mmWave, such as sensitivity to high co-channel interference, the lack of line of sight and the low propagation and penetration of the signal in outdoor environments. In these networks, links may suddenly disappear or their throughput may suddenly change. Therefore, the sudden and dynamic change in the throughput and delay will require a close interaction with the SDN controller, and determine not only new global paths, but also incorporating link stability in the choice of these paths.

To achieve the resiliency and agility needed at the mmWave backhaul, our approach is to collect additional metrics about the mmWave mesh network and link properties (i.e., packet loss, delay, RSSI, LQI, antenna direction \cite{mmwave}), in order to ensure faster recovery (in the presence of link failure), and optimized service selection. This information will be available to the SDN controller, and will be used to compute the link rates, in order to control the traffic flows inside the mesh network, optimizing the entire transport capacity of the mesh network between application services. With the metrics collected about the mmWave link state, the controller will be able not only to react to link failures, but also to predict before a link failure occurs, reconfiguring all the flows and forwarding tables according with the needs of the application services.


mmWave technology, on the other side, provides a great diversity on the establishment of different paths, achieving a multitude of end-to-end paths, and bringing multipath diversity to  the network that is ready to be explored by the application of MP-MH AC-RLNC techniques. The configuration of the mmWave mesh can therefore be self-organizedly performed to accommodate the users and services needs over time, and optimize the overall network conditions, being it higher throughput on the links or optimal balancing to improve delay and service performance metrics.

\subsection{Distributed storage and computation}

Caching and distributed storage using network coding are demonstrated to be very effective in exploiting all available network resources and reducing the peak traffic overload of central units \cite{maddah2014fundamental,dimakis2011survey}. Combining the caching and storage techniques, deployed in practice in today's modern networks, with the proposed adaptive and causal code proposed in this article, can satisfy the requirements of new applications that require ultra-reliable low-latency communications. As another enabling technology, we propose combining the idea of efficient distributed computation, and the adaptive and casual networking coding which results in ultra-reliable and low-delay distributed computations over a network of interconnected computational nodes.

A centralized computation scheme requires all data to be transmitted to a central unit for computation. However, with the enormous rate of data growth, the size of computational problems becomes larger and larger which results in huge communication and storage costs. Thus, the central computation scheme does not seem practical for large computational problems, e.g., multiplication of two large matrices, training a deep neural network, etc, and has led itself the viable and novel idea of distributed computation. In a distributed computation scheme, computations are performed at different nodes in the network which results in avoiding massive central computation, central storage, and unnecessary transmission of big chunks of data through a large parts of the network. In fact, computations at intermediate nodes can significantly reduce the communication and storage resource usage, and also reduce the process time per computational node. Distributed Computation has recently attracted significant attention. Recent interesting examples are, but not limited to, coded matrix multiplication \cite{sheth2018application,dutta2019optimal}, where the multiplication of two large matrices are split to several small matrix multiplication tasks over computational nodes of a simple network topology; and \cite{malak2020distribute} where a comprehensive information theoretical analysis is performed for achievable rates for distributed computation using network coding over a network with multiple sources and multiple destinations.

As an interesting envisioning technology, we propose combining the idea of distributed computation and the adaptive and casual networking coding which results in ultra-reliable and low-delay distributed computations over a network of interconnected computational nodes with unreliable communication links. The new MS-MD highly-meshed network architecture presented in this article provides the opportunity of performing  sequence of computational jobs defined in-order at source nodes to be distributed among  intermediate computational nodes. Then,  each receiver aggregates enough chunks of computational results  to identify the final computational results of the desired jobs in order. By borrowing introduced tools and ideas, e.g., local and global feedback, adaptation to the network changing dynamics, considering throughput-delay trade-off, etc, one can propose an effective adaptive and casual network coded computation over practical MS-MD networks.

\section{Acknowledgments}
The project SNOB-5G: Scalable Network Backhauling for 5G (reference CENTRO-01-0247-FEDER-045929) leading to this work is co-financed by the ERDF - European Regional Development Fund through the Operational Program for Competitiveness and Internationalisation - COMPETE 2020, the Center Portugal Regional Operational Program - CENTRO 2020 and the Portuguese Foundation for Science and Technology - FCT under the MIT Portugal Program.

\bibliographystyle{IEEEtran}
\bibliography{IEEEabrv,references}

\begin{thebibliography}{10}
\providecommand{\url}[1]{#1}
\csname url@samestyle\endcsname
\providecommand{\newblock}{\relax}
\providecommand{\bibinfo}[2]{#2}
\providecommand{\BIBentrySTDinterwordspacing}{\spaceskip=0pt\relax}
\providecommand{\BIBentryALTinterwordstretchfactor}{4}
\providecommand{\BIBentryALTinterwordspacing}{\spaceskip=\fontdimen2\font plus
\BIBentryALTinterwordstretchfactor\fontdimen3\font minus
  \fontdimen4\font\relax}
\providecommand{\BIBforeignlanguage}[2]{{%
\expandafter\ifx\csname l@#1\endcsname\relax
\typeout{** WARNING: IEEEtran.bst: No hyphenation pattern has been}%
\typeout{** loaded for the language `#1'. Using the pattern for}%
\typeout{** the default language instead.}%
\else
\language=\csname l@#1\endcsname
\fi
#2}}
\providecommand{\BIBdecl}{\relax}
\BIBdecl

\bibitem{Rezende2019}
{Rezende, Pedro and Kianpisheh, Somayeh and Glitho, Roch and Madeira, Edmundo},
  ``{An SDN-Based Framework for Routing Multi-Streams Transport Traffic Over
  Multipath Networks},'' in \emph{IEEE Int. Conf. Commun. (ICC)}.\hskip 1em
  plus 0.5em minus 0.4em\relax IEEE, may 2019, pp. 1--6.

\bibitem{bib:pi2016mmwave}
Z.~{Pi}, J.~{Choi}, and R.~{Heath}, ``{Millimeter-wave gigabit broadband
  evolution toward 5G: fixed access and backhaul},'' \emph{IEEE Communications
  Magazine}, vol.~54, no.~4, pp. 138--144, 2016.

\bibitem{noauthor_snob-5g_nodate}
\BIBentryALTinterwordspacing
``\BIBforeignlanguage{en-US}{{SNOB}-{5G} {Scalable} {Network} {Backhauling} for
  {5G}}.'' [Online]. Available: \url{https://snob-5g.com/}
\BIBentrySTDinterwordspacing

\bibitem{kahn1974protocol}
R.~Kahn and V.~Cerf, ``A protocol for packet network intercommunication,''
  \emph{IEEE Transactions on Communications}, vol.~22, no.~5, pp. 637--648,
  1974.

\bibitem{cerf1974specification}
V.~Cerf, Y.~Dalal, and C.~Sunshine, ``Specification of internet transmission
  control program,'' RFC 675, December, Tech. Rep., 1974.

\bibitem{kozierok2005tcp}
C.~M. Kozierok, \emph{{The TCP/IP guide: a comprehensive, illustrated Internet
  protocols reference}}.\hskip 1em plus 0.5em minus 0.4em\relax No Starch
  Press, 2005.

\bibitem{fall2011tcp}
K.~R. Fall and W.~R. Stevens, \emph{TCP/IP illustrated, volume 1: The
  protocols}.\hskip 1em plus 0.5em minus 0.4em\relax addison-Wesley, 2011.

\bibitem{martin2002tcp}
J.-P. Martin-Flatin and S.~Ravot, ``Tcp congestion control in fast
  long-distance networks,'' \emph{California Inst. Technol., Pasadena, CA, USA,
  Tech. Rep. CALT-68-2398}, 2002.

\bibitem{johansson2016congestion}
I.~Johansson, ``Congestion control for 4g and 5g access,'' \emph{Internet
  Engineering Task Force, Internet-Draft draft-johansson-cc-for-4g-5g-02},
  2016.

\bibitem{paasch2014multipath}
C.~Paasch and O.~Bonaventure, ``Multipath tcp,'' \emph{Communications of the
  ACM}, vol.~57, no.~4, pp. 51--57, 2014.

\bibitem{yedugundla2016multi}
K.~Yedugundla, S.~Ferlin, T.~Dreibholz, {\"O}.~Alay, N.~Kuhn, P.~Hurtig, and
  A.~Brunstrom, ``Is multi-path transport suitable for latency sensitive
  traffic?'' \emph{Computer Networks}, vol. 105, pp. 1--21, 2016.

\bibitem{ferlin2016blest}
S.~Ferlin, {\"O}.~Alay, O.~Mehani, and R.~Boreli, ``{BLEST}: {B}locking
  estimation-based {MPTCP} scheduler for heterogeneous networks,'' in
  \emph{2016 IFIP Networking Conference (IFIP Networking) and Workshops}.\hskip
  1em plus 0.5em minus 0.4em\relax IEEE, 2016, pp. 431--439.

\bibitem{ho2006random}
T.~Ho, M.~M{\'e}dard, R.~Koetter, D.~R. Karger, M.~Effros, J.~Shi, and
  B.~Leong, ``A random linear network coding approach to multicast,''
  \emph{IEEE Trans. Inf. Theory}, vol.~52, no.~10, pp. 4413--4430, 2006.

\bibitem{schneuwly2020discrete}
A.~Schneuwly, D.~Malak, and M.~M{\'e}dard, ``{D}iscrete water filling
  multi-path packet scheduling,'' in \emph{2020 IEEE International Symposium on
  Information Theory (ISIT)}.\hskip 1em plus 0.5em minus 0.4em\relax IEEE,
  2020, pp. 1658--1663.

\bibitem{chou2003practical}
P.~A. Chou, Y.~Wu, and K.~Jain, ``Practical network coding,'' in
  \emph{Proceedings of the annual Allerton conference on communication control
  and computing}, vol.~41, no.~1.\hskip 1em plus 0.5em minus 0.4em\relax The
  University; 1998, 2003, pp. 40--49.

\bibitem{patterson2014and}
S.~Patterson, ``How {MIT and Caltech's} coding breakthrough could accelerate
  mobile network speeds’,'' \emph{Network World}, 2014.

\bibitem{luby1997practical}
M.~G. Luby, M.~Mitzenmacher, M.~A. Shokrollahi, D.~A. Spielman, and V.~Stemann,
  ``Practical loss-resilient codes,'' in \emph{Proceedings of the twenty-ninth
  annual ACM symposium on Theory of computing}, 1997, pp. 150--159.

\bibitem{shokrollahi2006raptor}
A.~Shokrollahi, ``{R}aptor codes,'' \emph{IEEE/ACM Transactions on Networking
  (TON)}, vol.~14, no.~SI, pp. 2551--2567, 2006.

\bibitem{luby2002lt}
M.~Luby, ``{LT} codes,'' in \emph{Proc. IEEE Symp. Found. Computer
  Science}.\hskip 1em plus 0.5em minus 0.4em\relax IEEE, 2002, pp. 271--280.

\bibitem{GuoShiCaiMed2013}
W.~Guo, X.~Shi, N.~Cai, and M.~M\'{e}dard, ``{Localized dimension growth: A
  convolutional random network coding approach to managing memory and decoding
  delay},'' \emph{IEEE Trans. Commun.}, vol.~61, no.~9, pp. 3894--3905, Sep.
  2013.

\bibitem{ferlin2018mptcp}
S.~Ferlin, S.~Kucera, H.~Claussen, and {\"O}.~Alay, ``{MPTCP} meets {FEC}:
  Supporting latency-sensitive applications over heterogeneous networks,''
  \emph{IEEE/ACM Transactions on Networking}, vol.~26, no.~5, pp. 2005--2018,
  2018.

\bibitem{badr2016layered}
A.~Badr, P.~Patil, A.~Khisti, W.-T. Tan, and J.~Apostolopoulos, ``Layered
  constructions for low-delay streaming codes,'' \emph{IEEE Transactions on
  Information Theory}, vol.~63, no.~1, pp. 111--141, 2016.

\bibitem{fong2019low}
S.~L. Fong, S.~Emara, B.~Li, A.~Khisti, W.-T. Tan, X.~Zhu, and
  J.~Apostolopoulos, ``Low-latency network-adaptive error control for
  interactive streaming,'' in \emph{Proceedings of the 27th ACM International
  Conference on Multimedia}, 2019, pp. 438--446.

\bibitem{cloud2015coded}
J.~Cloud, D.~Leith, and M.~M{\'e}dard, ``A coded generalization of selective
  repeat {ARQ},'' in \emph{2015 IEEE Conf. Computer Commun}, 2015, pp.
  2155--2163.

\bibitem{sundararajan2011network}
J.~K. Sundararajan, D.~Shah, M.~M{\'e}dard, S.~Jakubczak, M.~Mitzenmacher, and
  J.~Barros, ``{N}etwork coding meets {TCP}: {T}heory and implementation,''
  \emph{Proceedings of the IEEE}, vol.~99, no.~3, pp. 490--512, 2011.

\bibitem{kim2012network}
M.~Kim, J.~Cloud, A.~ParandehGheibi, L.~Urbina, K.~Fouli, D.~Leith, and
  M.~M{\'e}dard, ``{Network coded TCP (CTCP)},'' \emph{arXiv preprint
  arXiv:1212.2291}, 2012.

\bibitem{cloud2013multi}
J.~Cloud, F.~du~Pin~Calmon, W.~Zeng, G.~Pau, L.~M. Zeger, and M.~Medard,
  ``Multi-path {TCP} with network coding for mobile devices in heterogeneous
  networks,'' in \emph{2013 IEEE 78th Vehicular Technology Conference (VTC
  Fall)}.\hskip 1em plus 0.5em minus 0.4em\relax IEEE, 2013, pp. 1--5.

\bibitem{Lin2019}
{Lin, Ying Dar and Liu, Te Lung and Wang, Shun Hsien and Lai, Yuan Cheng},
  ``{Proactive multipath routing with a predictive mechanism in
  software-defined networks},'' \emph{International Journal of Communication
  Systems}, vol.~32, no.~14, pp. 1--16, 2019.

\bibitem{langley2017quic}
A.~Langley, A.~Riddoch, A.~Wilk, A.~Vicente, C.~Krasic, D.~Zhang, F.~Yang,
  F.~Kouranov, I.~Swett, J.~Iyengar \emph{et~al.}, ``The quic transport
  protocol: Design and internet-scale deployment,'' in \emph{Proceedings of the
  Conference of the ACM Special Interest Group on Data Communication}, 2017,
  pp. 183--196.

\bibitem{de2019pluginizing}
Q.~De~Coninck, F.~Michel, M.~Piraux, F.~Rochet, T.~Given-Wilson, A.~Legay,
  O.~Pereira, and O.~Bonaventure, ``Pluginizing quic,'' in \emph{Proceedings of
  the ACM Special Interest Group on Data Communication}, 2019, pp. 59--74.

\bibitem{Pang2017}
{Pang, Junjie and Xu, Gaochao and Fu, Xiaodong}, ``{SDN-Based Data Center
  Networking With Collaboration of Multipath TCP and Segment Routing},''
  \emph{IEEE Access}, vol.~5, pp. 9764--9773, 2017.

\bibitem{cohen2020adaptive}
A.~Cohen, D.~Malak, V.~B. Brachay, and M.~Medard, ``Adaptive causal network
  coding with feedback,'' \emph{IEEE Trans Commun.}, 2020.

\bibitem{cohen2020adaptiveMPMH}
A.~Cohen, G.~Thiran, V.~B. Bracha, and M.~M{\'e}dard, ``Adaptive causal network
  coding with feedback for multipath multi-hop communications,'' in \emph{IEEE
  Int. Conf. Commun. (ICC)}.\hskip 1em plus 0.5em minus 0.4em\relax IEEE, 2020,
  pp. 1--7. Under minor revision IEEE Trans Commun. (arXiv preprint
  arXiv:1910.13\,290).

\bibitem{malak2019tiny}
D.~Malak, M.~M{\'e}dard, and E.~M. Yeh, ``{Tiny Codes for Guaranteeable
  Delay},'' \emph{IEEE J. Sel. Areas in Commun.}, vol.~37, pp. 809--825, 2019.

\bibitem{Xie2019}
{Xie, Junjie and Guo, Deke and Qian, Chen and Liu, Lei and Ren, Bangbang and
  Chen, Honghui}, ``{Validation of Distributed SDN Control Plane Under
  Uncertain Failures},'' \emph{IEEE/ACM Trans. Netw}, vol.~27, no.~3, pp.
  1234--1247, jun 2019.

\bibitem{Mamushiane2018}
{Mamushiane, Lusani and Lysko, Albert and Dlamini, Sabelo}, ``{A comparative
  evaluation of the performance of popular SDN controllers},'' \emph{IFIP
  Wireless Days}, vol. 2018-April, pp. 54--59, 2018.

\bibitem{Amin2018}
{Amin, Rashid and Reisslein, Martin and Shah, Nadir}, ``{Hybrid SDN networks: A
  survey of existing approaches},'' \emph{IEEE Communications Surveys and
  Tutorials}, vol.~20, no.~4, pp. 3259--3306, 2018.

\bibitem{Bannour2018}
{Bannour, Fetia and Souihi, Sami and Mellouk, Abdelhamid}, ``{Distributed SDN
  Control : Survey , Taxonomy , and Challenges},'' vol.~20, no.~1, pp.
  333--354, 2018.

\bibitem{Zhu2019}
\BIBentryALTinterwordspacing
{Zhu, Liehuang and Karim, Md Monjurul and Sharif, Kashif and Li, Fan and Du,
  Xiaojiang and Guizani, Mohsen}, ``{SDN Controllers: Benchmarking \&
  Performance Evaluation},'' pp. 1--14, Feb 2019. [Online]. Available:
  \url{http://arxiv.org/abs/1902.04491}
\BIBentrySTDinterwordspacing

\bibitem{Badotra2019}
{Badotra, Sumit and Panda, Surya Narayan}, ``{Evaluation and comparison of
  OpenDayLight and open networking operating system in software-defined
  networking},'' \emph{Cluster Computing}, vol.~0, 2019.

\bibitem{Secci2019}
{Secci, Stefano and Diamanti, Alessio and {Manuel Vilchez Sanchez}, Jos{\'{e}}
  and {Tahirou Bah}, Mamadou and Vizarreta, Petra and {Mas Machuca}, Carmen and
  Scott-Hayward, Sandra and Smith, Dylan}, ``{Security and Performance
  Comparison of ONOS and ODL controllers},'' ONF, Tech. Rep., 2019.

\bibitem{Liu2018}
{Liu, Jed and Hallahan, William and Schlesinger, Cole and Sharif, Milad and
  Lee, Jeongkeun and Soul{\'{e}}, Robert and Wang, Han and Caşcaval, Călin
  and McKeown, Nick and Foster, Nate}, ``{p4v: Practical Verification for
  Programmable Data Planes},'' in \emph{Proceedings of the 2018 Conference of
  the ACM Special Interest Group on Data Communication}.\hskip 1em plus 0.5em
  minus 0.4em\relax ACM, aug 2018, pp. 490--503.

\bibitem{5growth_d22}
H.~G. Project, ``{D2.2 : Initial implementation of 5G End-to-End Service
  Platform},'' Tech. Rep., 2020.

\bibitem{Guan2019}
{Guan, Bowei and Shen, Shan Hsiang}, ``{FlowSpy: An efficient network
  monitoring framework using P4 in software-defined networks},'' \emph{IEEE
  Vehicular Technology Conference}, vol. 2019-September, 2019.

\bibitem{p4Lang}
{The P4 Language Consortium}, ``{P4 Language Specification},''
  https://p4.org/p4-spec/docs/P4-16-v1.2.1.html, 2020.

\bibitem{stratum}
{Open Network Foundation}, ``{Stratum: Enabling the era of next generation
  SDN},'' https://www.opennetworking.org/stratum/, 2020.

\bibitem{onos_apps_kafka}
{ONF}, ``{}kafka-onos,'' https://github.com/opencord/kafka-onos, 2020.

\bibitem{Janz2016}
{Janz, Christopher and Ong, Lyndon and Sethuraman, Karthik and Shukla, Vishnu},
  ``{Emerging transport SDN architecture and use cases},'' \emph{IEEE
  Communications Magazine}, vol.~54, no.~10, pp. 116--121, 2016.

\bibitem{ONF_TAPI}
{Open Network Foundation}, ``{TAPI v2.1.3 Reference Implementation Agreement
  TR-547, Version 1.0},'' July 2020.

\bibitem{pedersen2011kodo}
M.~V. Pedersen, J.~Heide, and F.~H. Fitzek, ``{Kodo: An open and research
  oriented network coding library},'' in \emph{International Conference on
  Research in Networking}.\hskip 1em plus 0.5em minus 0.4em\relax Springer,
  2011, pp. 145--152.

\bibitem{PRNG}
\BIBentryALTinterwordspacing
``\BIBforeignlanguage{en-US}{{S}liding {W}indow {R}andom {L}inear code ({RLC})
  {F}orward {E}rasure {C}orrection ({FEC}) {S}chemes for {FECFRAME}}.''
  [Online]. Available: \url{https://tools.ietf.org/html/rfc8681}
\BIBentrySTDinterwordspacing

\bibitem{PRNG_Generator}
\BIBentryALTinterwordspacing
``\BIBforeignlanguage{en-US}{{T}iny{MT32} {P}seudorandom {N}umber {G}enerator
  ({PRNG})}.'' [Online]. Available: \url{https://tools.ietf.org/html/rfc8682}
\BIBentrySTDinterwordspacing

\bibitem{quic_fec_sof}
\BIBentryALTinterwordspacing
``\BIBforeignlanguage{en-US}{{S}liding {W}indow {R}andom {L}inear {C}ode
  ({RLC}) {F}orward {E}rasure {C}orrection ({FEC}) {S}chemes for {QUIC}}.''
  [Online]. Available:
  \url{https://tools.ietf.org/html/draft-roca-nwcrg-rlc-fec-scheme-for-quic-03\#section-4.1.3}
\BIBentrySTDinterwordspacing

\bibitem{sallam2019security}
A.~Sallam, A.~Refaey, and A.~Shami, ``{On the Security of SDN: A Completed
  Secure and Scalable Framework Using the Software-Defined Perimeter},''
  \emph{IEEE Access}, vol.~7, pp. 146\,577--146\,587, September 2019.

\bibitem{lee2020comprehensive}
S.~Lee, J.~Kim, S.~Woo, C.~Yoon, S.~Scott-Hayward, V.~Yegneswaran, P.~Porras,
  and S.~Shin, ``A comprehensive security assessment framework for
  software-defined networks,'' \emph{Computers \& Security}, vol.~91, April
  2020.

\bibitem{cohen2020PostQuant}
A.~Cohen, R.~D’Oliveira, S.~Salamatian, and M.~M{\'e}dard, ``{Network
  Coding-Based Post-Quantum Cryptography},'' \emph{arXiv preprint
  arXiv:2009.01931}, 2020.

\bibitem{mmwave}
{M. Agiwal and A. Roy and N. Saxena}, ``{Next generation 5G wireless networks:
  A comprehensive survey},'' \emph{IEEE Communication Surveys Tutorials},
  vol.~18, no.~3, pp. 1617--1655, jun 2016.

\bibitem{maddah2014fundamental}
M.~A. Maddah-Ali and U.~Niesen, ``Fundamental limits of caching,'' \emph{IEEE
  Transactions on Information Theory}, vol.~60, no.~5, pp. 2856--2867, 2014.

\bibitem{dimakis2011survey}
A.~G. Dimakis, K.~Ramchandran, Y.~Wu, and C.~Suh, ``A survey on network codes
  for distributed storage,'' \emph{Proceedings of the IEEE}, vol.~99, no.~3,
  pp. 476--489, 2011.

\bibitem{sheth2018application}
U.~Sheth, S.~Dutta, M.~Chaudhari, H.~Jeong, Y.~Yang, J.~Kohonen, T.~Roos, and
  P.~Grover, ``An application of storage-optimal matdot codes for coded matrix
  multiplication: Fast k-nearest neighbors estimation,'' in \emph{2018 IEEE
  International Conference on Big Data (Big Data)}.\hskip 1em plus 0.5em minus
  0.4em\relax IEEE, 2018, pp. 1113--1120.

\bibitem{dutta2019optimal}
S.~Dutta, M.~Fahim, F.~Haddadpour, H.~Jeong, V.~Cadambe, and P.~Grover, ``On
  the optimal recovery threshold of coded matrix multiplication,'' \emph{IEEE
  Transactions on Information Theory}, vol.~66, no.~1, pp. 278--301, 2019.

\bibitem{malak2020distribute}
D.~Malak, A.~Cohen, and M.~M{\'e}dard, ``How to distribute computation in
  networks,'' in \emph{IEEE INFOCOM 2020-IEEE Conference on Computer
  Communications}.\hskip 1em plus 0.5em minus 0.4em\relax IEEE, 2020, pp.
  327--336.

\end{thebibliography}

\end{document}